\documentclass{elsart}
\usepackage{harvard}
\usepackage{amssymb,epsf}
\usepackage{graphicx}
\usepackage{color}

\makeatletter
\def\elsartstyle{%
	\def\normalsize{\@setfontsize\normalsize\@xiipt{14.5}}
	\def\small{\@setfontsize\small\@xipt{13.6}}
	\let\footnotesize=\small
	\def\large{\@setfontsize\large\@xivpt{18}}
	\def\Large{\@setfontsize\Large\@xviipt{22}}
	\skip\@mpfootins = 18\p@ \@plus 2\p@
	\normalsize
}
\makeatother

\def\url#1{{\ttfamily\def\/{/\discretionary{}{}{}}#1}}

\newcommand{\be}{\begin{equation}}
\newcommand{\ee}{\end{equation}}
\def\bea{\begin{eqnarray}}
\def\eea{\end{eqnarray}}
\newcommand\tps{temperatures~}  
\newcommand\tew{emission-weighted~} 
\newcommand\tmw{$T_{mw}^{200}$~}

\newcommand{\msu}{h^{-1} M_{\odot}}  

\newcommand{\La}{{$\Lambda$CDM~}}
\newcommand{\rs}{r/r_{200}}

\newcommand{\gr}{\kern 2pt\hbox{}^\circ{\kern -2pt K}} 
\newcommand{\oml}{\Omega_{\Lambda}}
\newcommand{\omm}{\Omega_{m}}
\newcommand{\brr}{\begin{array}}
\newcommand{\err}{\end{array}}
\newcommand{\ltsima}{$\; \buildrel < \over \sim \;$}
\newcommand{\simlt}{\lower.5ex\hbox{\ltsima}}
\newcommand{\gtsima}{$\; \buildrel > \over \sim \;$}
\newcommand{\simgt}{\lower.5ex\hbox{\gtsima}}

\begin{document}
\begin{frontmatter}

\title{ X-ray temperature spectroscopy of simulated cooling clusters} 

\author{{\thanksref {valda}} Riccardo Valdarnini}
\thanks[valda]{E--mail: valda@sissa.it}

\address{SISSA -- International School for Advanced Studies,
    Via Beirut 2/4, Trieste, Italy}

{\it Corresponding author} : R. Valdarnini

\begin{abstract}
Results from a large sample of hydrodynamical/N-body simulations of galaxy 
clusters in a \La cosmology are used to simulate cluster X-ray observations. 
The physical modeling of the gas includes radiative cooling, star formation,
energy feedback and metal enrichment that follow from supernova  explosions.
Mock cluster samples are constructed grouping simulation data according to 
a number of constraints which would be satisfied by a data set of
 X-ray measurements of cluster \tps as expected from {\it Chandra} 
observations. 
The X-ray spectra from simulated clusters are fitted into different 
energy bands using the XSPEC {\it mekal} model.
The biasing of spectral \tps with respect to mass-weighted \tps 
 is found to be influenced by two independent processes.
The first scale dependency is absent 
in adiabatic runs and is due to cooling, whose efficiency to transform 
cold gas into stars is higher for cool clusters and this in turn implies
a strong dependency of the spectral versus mass-weighted 
temperature relation on the cluster mass. The second dependency is due to 
photon 
emission because of cool gas which is accreted during merging events and 
biases the spectral fits. These events have been quantified according to the 
power ratio method and a robust correlation is found to exist 
between the spectral bias and the amount of cluster substructure. 

The shape of the simulated temperature profiles is not 
universal and it is steeper at the cluster center for cool clusters than for
the massive ones. This follows owing to the scale dependency introduced by 
cooling which implies for cool clusters higher central temperatures, in 
scaled units, than for massive clusters.
The profiles are in good agreement with data in the radial range 
between $\sim 0.1 r_{vir}$ and $\sim 0.4 r_{vir}$; at small radii 
($r\simlt 0.1 r_{vir}$) the cooling runs fail to reproduce the shape of 
the observed profiles. The fit is improved if one considers a hierarchical 
merging scenario in which cluster cores can accrete cooler gas through 
merging with cluster subclumps, though the shape of the temperature 
profiles is modified in a significant way only in the regime where 
the mass of the substructure is a large fraction of the cluster mass.

\end{abstract}

\end{frontmatter}
%

\section{Introduction}
\label{sec:introduction}

Observations show that most of the baryons present in a galaxy cluster are in the form of a 
hot ionized gas (intracluster medium , ICM), with temperatures in the 
range $\sim 10^6 -10^7 \gr$ \cite{sar86}.
The continuum emission is dominated by
bremsstrahlung processes, and is free from the contamination effects which
 may arise in the optical band. For this reason, observations of galaxy 
clusters in the X-ray band have been extensively carried out over the past 
decade \cite{hen92,ebe97,ros98}
X-ray observations of the spatial distribution of cluster density and 
temperature  allow one to determine, under the assumption of hydrostatic
equilibrium, the mass in baryons as well as the dark matter profile and 
the total cluster mass $M_X$. The latter quantity is connected to the 
global cluster temperature $T_X$, and through 
the $M_X-T_X$ relation the cluster mass function can then be used to find the cluster X-ray
temperature or luminosity functions. Observations of these quantities are 
thus used to constrain theoretical models 
\cite{hen91,edg90,hen97,ros98,ebe98}. 
Knowledge of the temperature profile is also important in order to assess
 the role of non-gravitational heating processes which can contribute to the
 budget of the gas thermal energy. This is relevant because observations 
 show that the cluster scaling relations do not obey the self-similar 
behavior predicted by gravitational collapse \cite{dav93,ala98,mar98}. 
Heating of the ICM by non-gravitational processes will break self-similarity 
 and help to explain 
the observed relations \cite{evr91,kai91,whi91,low96,va99,lo01}.

A powerful tool for studying the dependence of cluster observational
quantities upon theoretical models is given by the use of hydro/N-body
simulations. Hydrodynamic simulations have been widely used to investigate 
the formation and evolution of galaxy clusters
\cite{kat93,sug98,ann96,yos00,per20,lew20,bem01,mu02,lo02,va03,tor03,as03},
see also Voit (2005) and references cited therein.
The advantage over analytical methods is that they can treat 
the gas evolution self-consistently.
The validity of the numerical approach is supported by X-ray observations
of surface brightness maps, which indicate the existence of merging 
substructure \cite{mo95}. Moreover, high resolution observations performed 
with {\it Chandra} and XMM-Newton satellites have revealed  the existence of 
complex cluster temperature structure, like shocks induced by mergers  
and 'cold front' \cite{ma00,vi01,ma02}.

Hydrodynamic simulations in which modeling of the gas physical processes 
incorporates radiative cooling, star formation and energy feedback from 
supernovae yield a cluster luminosity-temperature relation in good
agreement with data for cluster temperatures $T_X \simgt 1 keV$ 
\cite{va03,bo04}.
These results support the radiative cooling model \cite{br00}, 
where cluster X-ray properties are determined by the cooling 
efficiency to turn cold gas into stars.
However, there are a number of issues for which hydrodynamic simulations
that include the physical modeling of the gas described above do not 
produce a satisfactory agreement with data.
For instance, the radial temperature profiles are not isothermal and decline 
with radius.
At small distances from the cluster center the 
simulated profiles have a steep rise moving inwards.
This behavior is not seen in a set of observations 
\cite{ma98,al01,de02,vi05,pi05}.
The data agree with the simulation results at $r \simgt 20\% r_{vir}$ 
but indicate a temperature profile which is flat or declining at 
very small distances.  
These data are also at variance  with the predictions of the 
standard cooling flow model \cite{fa94}. In the very central cluster regions
the amount of cool gas is smaller than that predicted by the model,
with gas temperatures never below $\simeq 1 keV$ \cite{ka01,ta01,pe01}.
 This failure to 
reproduce temperature data has prompted many authors to consider possible
heating mechanisms to explain the lack of cool gas in the cluster core.
Among the proposed models, thermal conduction \cite{voi02,jub04,do04,si06} and 
energy feedback
 from  active galactic nuclei \cite{cu01,br01,bo02,fab02,br02,om04,re05} 
are the most considered.

The observed temperature profiles are then sensitive to different physical
phenomena and can be used to constrain theoretical models. Moreover, as
already outlined, the intracluster gas exhibits a complex thermal structure.
It is then important to perform properly the comparison of X-ray 
observations 
with the results of numerical simulations, taking into account the effects of
instrumental response and background contaminations on the    
 spectral fit temperatures. As an additional effect which can bias  
comparisons, the spectral temperatures are measured 
in a specified energy bands, whereas the simulated cluster temperatures 
are theoretically defined according to a specified weighting scheme.
This problem has been already analyzed by Mathiesen \& Evrard (2001) and, 
more recently, by Gardini et al. (2004), Mazzotta et al. (2004), Rasia et al. 2005
 and Vikhlinin  (2006).
The aim of this paper is to investigate for systematic effects which 
can bias spectroscopic measurements of cluster X-ray temperatures.
For this purpose, a large numerical sample of simulated clusters is used to
 construct 
spatially resolved X-ray spectra as expected from observations
with the {\it Chandra} satellite.
The cluster spectral fit temperatures are found using the XSPEC library,
and are compared against mean gas temperatures with theoretically defined
averages. 
Moreover, spatially resolved X-ray spectra of the simulated clusters are also
 used to
investigate how the measured temperature profiles differ from the projected
profiles obtained directly from simulations.
The comparisons are performed on a statistical basis, using the  
cluster numerical sample.
The numerical sample is also subdivided according to the amount of 
substructure 
present in a given cluster, this allows us to investigate how spectral 
measurements are affected by the cluster dynamical state.

This paper can be considered a generalization of that of Mathiesen \& Evrard 
(2001, hereafter ME), the main 
differences being the size of the numerical cluster sample used, together
 with a more complete physical modeling of the gas in the  
simulations and the X-ray spectroscopic analysis of radial temperature 
profiles.
The paper is organized as follows. The simulations are presented in Section 
2, in Section 3 the procedures with which average 
and spectroscopic temperatures are extracted from the simulated clusters
are discussed.
Section 4 is dedicated to the method used to measure substructure in the
simulated cluster. The results are discussed in Section 5 and the conclusions
are drawn in Section 6.

\section{Simulations} \label {sec:simulations}

The simulations were run using a TREESPH code. A detailed description of 
the simulation procedure is given in Valdarnini (2003, hereafter V03)
The cosmological model assumes a flat CDM universe, with 
matter density parameter $\omm=0.3$, $\oml=0.7$, $\Omega_b=0.019h^{-2}$ 
 and $h=0.7$ is the value of the Hubble constant in units of 
$100 Km sec^{-1} Mpc^{-1}$. 
The primeval spectral index of the power spectrum $n$ is set to $1$ and 
the power spectrum has been normalized to $\sigma_8=0.9$ on a 
$8h^{-1} Mpc$ scale.
Initial conditions for the cluster simulations are constructed as 
follows, following a two-step procedure.
A collisionless cosmological simulation is first run in a box of 
comoving size $L$, using a P3M code with $N_p$ particles, starting from an
initial redshift $z_i=10$.

 Clusters of galaxies were located at $z=0$ using a friends--of--friends 
(FoF) algorithm, so as to detect overdensities in excess of $\simeq 200
\Omega_m^{-0.6}$ within a radius $r_{vir}$.
The corresponding mass $M_{vir}$ contained within 
this radius is defined as 
 $M_{vir}= (4 \pi/3) \Omega_m \rho_c \Delta_c r_{vir}^3$,
where $\Delta_c =187 \Omega_m^{-0.55}$ for a flat cosmology and $\rho_c$ is the 
critical density.
A cluster numerical sample is constructed with the cluster identified using 
this procedure, with the clusters being sorted according to the values 
of their $M_{vir}$ at $z=0$.
The final numerical sample used in the hydrodynamic simulations consists 
of two distinct samples, S1 and S2.
The first sample  comprises the most massive $n_1=120$ clusters 
identified at $z=0$ in a cosmological run with $L_1=200h^{-1}Mpc$ and 
$N_{1p}=84^3$ particles.
Sample S2  consists of the $n_2$ most massive clusters identified in a new
cosmological run with box size $L_2=400h^{-1}Mpc$ and $N_{2p}=168^3$.
The number  $n_2(=32)$ of clusters in sample S2 is chosen so that the mass 
 $M_{vir}$ of the $n_2-th$ least massive  cluster is just above
that of the first cluster of the sample S1.
 For sample S2 $M_{vir}$ ranges from $\simeq 10^{15} \msu$ down to 
$2.3 \cdot 10^{14} \msu$, while for sample S1 the $n1-th$ cluster has 
 $M_{vir}= 10^{13} \msu$.
The random realization of the initial density perturbations are different
in the two cosmological simulations.
The final numerical cluster sample of $153$ clusters represents then a 
complete sample
of the clusters present in a cubic region of size $400 h^{-1}Mpc$
down to virial masses $\simeq 2.3\cdot 10^{14} \msu$, and is 
undersampled by  a factor $8$ below this mass and 
$M_{vir} \simeq 10^{13}\msu$.

Hydrodynamic simulations are performed in physical coordinates for each of the
clusters present in the sample. The initial conditions are constructed as
follows. 
All of the cluster particles at $z=0$ which are within a sphere of radius 
$r_{vir}$ located at the cluster center are identified.
These particles are traced back to a redshift $z_{in}=49$ and a cube of 
size $L_c \simeq 25-50 Mpc$ enclosing these particles is placed at the 
cluster center. 
A lattice of $N_L=51^3$ grid points is set inside the cube, and to each
node is associated a gas particle with its mass and position.
A similar lattice is set for dark matter particles, with the node positions
being displaced by one-half of the grid spacing with respect the nodes of
the gas lattice.
The particle positions are then perturbed, using the same random realization
as for the cosmological simulations. High-frequency modes are added to the 
original random realization in order to sample the increased Nyquist frequency.
The gas and dark matter particles used for the hydrodynamic simulations
are those for which the perturbed particle positions lie within a sphere
of radius $L_c/2$ from the cluster center.
To model external gravitational fields these particles are surrounded 
out to a radius $L_c$ by a spherical shell of low-resolution dark matter 
particles, each having a mass $8$ times the sum of the masses of a
gas and dark matter particle of the inner region.

The simulations are evolved to the present using a multistep TREESPH code
with a tolerance parameter $\theta=1$, quadrupole corrections enabled, and
minimum timesteps of $\Delta t_{min}= 6.9 \cdot 10^5 yr$ and 
$\Delta t= 8 \Delta t_{min}$ for gas and dark matter particles, respectively.
For the clusters of sample S2(S1), simulations have been performed setting
 $\varepsilon_g =25(15) kpc$, where $\varepsilon_g$  is the gravitational 
softening parameter of the gas particles.
The gravitational softening parameters of other species of 
particles have been set according 
 to the scaling $\varepsilon_i \propto m_i^{1/3}$, where $m_i$ is the mass 
of the particle $i$.
The runs  have a number of gas particles 
$N_g \simeq 70,000$, with similar values for the number of dark matter 
particles in the inner and outer shell. 
The mass of the gas particles ranges from $m_g \simeq 5.5 \cdot 10^9  M_{\odot}$
for the first cluster of sample S2, down to 
 $m_g \simeq 6 \cdot 10^8 M_{\odot}$ for the least massive cluster 
of sample S1.
For the runs considered here this mass resolution can be considered adequate
(Valdarnini 2002), yielding  final gas 
distributions with converging profiles for the simulated clusters.
 {This issue will be discussed in more detail in sect. 6.}

Numerical integrations have been performed with comoving softenings 
out to $z=20$, after which they are kept fixed in physical coordinates. 
Hydrodynamic simulations are followed according to the SPH method
(Hernquist \& Katz 1989), with a minimum smoothing length 
of $\varepsilon_g /4$.
During its evolution the thermal energy equation is subject to 
energy sinks from recombination and collisional excitation, bremsstrahlung and 
inverse Compton cooling. 
The cooling function $\Lambda_c(T,Z)$ depends on the gas temperature and 
metallicity.   
Lock-up tables have been constructed from Sutherland \& Dopita (1993)
and stored in a file. During the simulations, the tabulated values 
are interpolated to obtain $\Lambda_c(T,Z)$ for the gas particles.
Cold gas in high density regions is thermally unstable and is subject
to star formation (SF). A gas particle which is in a collapsing region and
whose cooling time is smaller than the dynamical time $\tau_d$ is eligible 
to form in a timestep $\Delta t$ a star particle with half of its mass, at 
 a rate given by the $\tau_d^{-1}$.
At each timestep a star particle heats its gas neighbors by 
supernova (SN) explosions of type II and Ia.
The number of SN explosions in the time interval is calculated according to 
the star particle age and the initial mass function, with each SN explosion
adding $ \simeq 10^{51} $ ergs to the gas thermal energy.
An Arimoto \& Yoshii (1987)  initial mass function has been assumed (V03),
 in the mass range from  $0.1$ to $40 M_{\odot}$.
The gas is also metal enriched at each timestep through SN explosions.
The mass in metals ejected by a star particle is distributed among 
its neighbors according to the SPH smoothing procedure.
The gas metallicity thus increases with time, and the dependence of the 
cooling rate on the gas metallicity is taken into account properly.
In order to construct simulated X-ray spectra from a given cluster all the 
relevant hydrodynamical variables of an individual run are output 
during the simulation at various redshifts, with the smaller ones 
being $z=0.025,0.039,0.052$.

\section{Temperature  definitions} \label {sec:temperature}
In this section the way in which mean cluster gas temperatures 
and temperature profiles with theoretically 
defined averages are obtained  from simulations data, 
together with the spectral temperatures, is described. 
The simulation temperatures are defined 
as those obtained from theoretically defined averages applied to the 
gas particle temperatures of output data. 
A spectroscopic X-ray measurement
of a galaxy cluster temperature is constructed with a much less simple 
procedure. 
Using output data produced with the runs described in Section 2,
X-ray spectra are first obtained from a certain cluster at a specified 
redshift.
As a second step a background component has been added to the spectra, 
and the resulting spectra have been then properly convolved with a template 
area response file (ARF)
and redistribution matrix file (RMF) of the {\it Chandra} observatory.
Finally, the generated event files are fitted by a single temperature 
{\it mekal}
model available with the XSPEC library.

\subsection{Simulation temperatures}

In the continuum limit an average gas temperature is defined as

\be
T_{\it W}^R= \frac {\int_0^R T(\vec x ) {\it W} d^3x } { \int_0^R {\it W} d^3x},
\ee
where $\it W$ is a weight function and the integral has a spherical boundary
of radius $R$. Common choices for $\it W$ are the gas 
density $\rho_g(\vec x)$ (mass-weighted temperatures, ${\it W}=\rho_g$) and 
the X-ray emissivity $\varepsilon^{(X)}( \vec x )= \Lambda^{(X)}(T,Z) 
n_en_H$ 
(emission-weighted temperatures, ${\it W}=\varepsilon^{(X)}$), $n_e$ and 
$n_H$ being
the electron and hydrogen gas number density, respectively.
The cooling function $\Lambda^{(X)}(T,Z)$ is calculated using a Raymond-Smith
 (1977) code.
Because of the Lagrangian nature of SPH simulations, volume integrals are 
replaced by a summation over the particles, so that
\be
T_{w}^R=\sum_i {\it W}_i \frac {m_i} {\rho_i} T_i / \sum_i {\it W}_i \frac 
{m_i}{\rho_i},
\label{tw}
\ee
where the subscript $i$ denotes the value of the quantity at the 
 position of the particle $i$, $\rho=\rho_g$,~$w=m$ or $w=ew$ according to 
the chosen weighting scheme and the summation is understood to be over all
 the gas
 particles within a distance  $R$ from the cluster center. {The cluster center is 
defined to be the position $\vec x_c$ where the gas 
density reaches its maximum. The position $\vec x_c$ is found according to the 
following iterative procedure: at each iteration $k=1,2,..,M$ the 
center of mass is found, $\vec x_c^{(k)}$, based on the gas particles which are 
within a sphere of radius $R^{(k)}<R^{(k-1)}$, located at $\vec x_c^{(k-1)}$.
The last iteration $M$ is such that there must be at least a number $N_M$ of 
gas particles within $R^{(M)}$. Robust results are obtained for 
$N_M \simeq 50-100$ 
and $R^{(k)}=(3/4)R^{(k-1)}$. The cluster center is then defined as 
$\vec x_c^{(M)}$.}

The emission-weighted temperature $T_{ew}^R$ is defined using the bolometric
 emissivity, in a similar way a band-limited emission-weighted temperature 
$T_{ew}^R(E_1-E_2)$ can be defined in the 
energy range $E_1-E_2$ by using the band-limited emissivity $\varepsilon=
\int_{E_1}^{E_2} \varepsilon_{\nu} d \nu$, where $\varepsilon_{\nu}$ is 
the specific emissivity.
It is useful to define a radius $R_{\Delta}$ such that
the average density within that radius is $\Delta$ times the critical density, 
i.e. $M_{\Delta}= {4 \pi R_{\Delta}^3} \Delta \rho_c(z) /3$, where
$\rho_c(z)=3 H(z)^2 /8 \pi G $,~$H(z)^2=H_0^2 E(z)^2$ and 
$E(z)^2=\Omega_m (1+z)^3 + \Omega_{\Lambda}$.
Hereafter, average global cluster temperatures will be denoted by 
$T_{w}^{\Delta}$, 
where $\Delta$ takes the values commonly used in  
literature, i.e. $\Delta=2500,~500$ and $200$.

Temperature profiles are obtained by first choosing a line of sight and then
locating in the plane orthogonal to the line of sight
annuli of increasing radii around the cluster center.
Hydrodynamic variables are estimated at a set of grid points using the SPH 
smoothing procedure. The geometry of the grid is cylindrical with coordinates
($\tilde {\rho}, \tilde{\phi} ,\tilde{z}$) and the origin is located at the 
cluster center, the $\tilde{z}-$axis is along
 the line of sight. The points are uniformly spaced, linearly in the angular
coordinate, and logarithmically in the radial and $\tilde{z}$ coordinates.
The range of the spatial coordinates is between $2 \cdot 10^{-4}$ and unity
in units of $R_{200}$.  
There are $N_{\tilde {\rho}}=60, N_{\tilde{\phi}}=20, 2N_{\tilde{z}}=50$ points
in the coordinate intervals.
For a continuous distribution the annulus having a finite width with 
projected radial boundaries $b_1$ and $b_2$ 
has a projected emission-weighted temperature $T(b)$ given by 
\be
T(b)= \int_{b_1}^{b_2} \tilde {\rho} d\tilde {\rho} 
\int_{-\infty}^{+\infty} \varepsilon^{(X)} T(\vec x ) d \tilde {z}
\left / \int_{b_1}^{b_2} \tilde {\rho} d\tilde {\rho} 
\int_{-\infty}^{+\infty} \varepsilon^{(X)} d \tilde {z} \right. ,
\label{tb}
\ee
where $b= (b_1+b_2)/2$  and $|\vec x|^2= \tilde {\rho}^2+\tilde {z}^2$.
The integrals (\ref{tb}) are approximated by first evaluating the 
corresponding hydro variables at the grid points and then
performing the discrete volume integrals over the set of
 points which are within the given boundaries, the integral 
along the $\tilde {z}-$axis is truncated at $|\tilde {z}|=R_{200}$.
The projected emission-weighted  temperatures are calculated at a set of 20
projected radii, with approximately uniform intervals in $\log b$ and 
ranging between $b_{min}\simeq 5 \cdot 10^{-3}$ and $b_{max}\simeq 0.5$ 
in units of $R_{200}$.

\subsection{Spectral temperatures}
The simulated spectroscopic temperatures are obtained by fitting the
cluster spectral emission with single temperature models.
The procedure to obtain spectroscopic \tps from simulated source spectra
consists of three separate steps which are described in what follows. 
The first step is to obtain photon spectra from a simulated cluster observed
 at a redshift $z$ along a given line of sight.
The  photon flux per unit energy $S_{\nu}$ (photons/sec cm$^2$ keV) that 
reaches the observer is given by 
\be
S_{\nu}= (1+z)^2 L_{\nu(1+z)}  /  4 \pi d_{L}^2(z),
\label{snu}
\ee
where  $d_L$ is the luminosity distance and $L_{\nu}$ is the volume integral 
of the specific X-ray emissivity of the cluster:
$L_{\nu}=\int \varepsilon_{\nu}^{(X)} d V.$ 
The cluster global spectral fit temperatures $T_s^{\Delta}$, to be 
compared with the averaged cluster 
temperature $T_w^{\Delta}$,  is obtained from
 a flux $S_{\nu}^{\Delta}$ with volume integral performed as in Eq. (\ref{tw}),
 over the same spherical boundaries.
The cluster spectral temperature profile $T_s(b)$ is found using 
  for each ring the photon flux $S_{\nu}(b)$. The fluxes  are calculated 
with the volume integrals over $\varepsilon^{(X)}_{\nu}$ 
in Eq. (\ref{snu})  being performed
in an analogous way to those over $\varepsilon^{(X)}$
in Eq. (\ref{tb}), using the same grid geometry and boundaries.

The second step consists of manipulating the simulated photon spectra 
$S_{\nu}$ according to 
a certain procedure, in order to reproduce the spectroscopic 
temperatures 
that would be obtained by fitting the cluster spectra as measured by 
 the {\it Chandra} Advanced Camera for Imaging and Spectroscopy (ACIS) 
instrument. In particular, the spectral analysis described here will 
consider observations relative to the S3 chip of the ACIS configuration.
The S3 chip has a small field of view ($\theta_{S3}=8.4 ~arc min$), but 
with a high spatial ( 1 pixel=$0.5~ arc sec$) and energy resolution
($\Delta E \simeq 100eV$).
For the considered energy band the simulated spectra are then binned
into energy channels with width $\Delta E=100 eV$, an energy resolution
which is adequate to correctly model the
 {\it Chandra} ACIS energy resolution (ME).
The cluster global spectral fit temperatures $T_s^{\Delta}$  
are calculated in two energy bands :
$[2-10]$ and $[0.5-10]keV$.
Spectroscopic temperature profiles $T_s(b)$ are instead considered in the 
energy band $[0.5-7]keV$. {Hereafter it is understood that all the considered 
energy bands are corrected for the cluster redshifts.}
Once the fluxes $S_{\nu}$ have been discretized the files which 
contain the spectral distribution $S_{\Delta E}$ are transformed into 
new files $S_{\Delta E}(XS)$, written in the FITS file format 
\footnote{http://fits.gsfc.nasa.gov/}.
 This  format allows the files
to be read by the XSPEC (v 11.3.0) library 
\footnote{http://heasarc.gsfc.nasa.gov/docs/xanadu/xspec/} used to construct
 the modeled spectra.

The background is modeled by adding  a  background  file $B_{\Delta E}(XS)$
to each spectral file  $S_{\Delta E}(XS)$.
The file $B_{\Delta E}(XS)$ with the proper spectral 
distribution of background photons is
constructed as follows. The normalization parameters are first 
calculated using, for different energy bands, the quiescent background rates 
$B_a$  (photons/sec arcmin$^2$) taken from 
Tables 1 and 2 of Markevitch 
\footnote{http://cxc.harvard.edu/contrib/maxim/bg/index.html}.
For a mock observation with exposure time $t_{exp}$, aperture angle
$\Delta \theta$ and a given energy band, the corresponding
background rate is used to calculate the expected number of background 
photons $N_B$. 

A file $B_{\Delta E}(XS)$ with a background spectrum is then created from a 
template background file 
\footnote{http://cxc.harvard.edu/cal/Acis/Cal\_prods/bkgrnd/current/} 
using the {\it fakeit} command in XSPEC. 
The spectrum is convolved 
with the appropriate ARFs and RMFs used to mimic the instrument responses 
(see Sect. 3.4), which are the same pair of matrices used to perform the 
convolution of the source spectrum  $S_{\Delta E}(XS)$.
The parameters of the background file are rescaled so that when
 integrating the spectrum over $t_{exp}$ seconds there
are $N_B$ photons in the energy band under consideration.
To this end the files are manipulated using the utility {\it fparkey}  from
the FTOOLS library  
\footnote{http://heasarc.gsfc.nasa.gov/lheasoft/ftools/}.
A companion file $B_{\Delta E}^{(mk)}(XS)$ is created for later use with 
the same procedure.
At this stage the binned source spectrum $S_{\Delta E}(XS)$ 
is convolved with the ARF and RMF files of the mock observation
using the {\it atable} tool, available with XSPEC.
The produced pulse height amplitude (PHA) file 
and the corresponding background file are added together
by applying to the files the {\it mathpha} utility of the FTOOLS library.
Errors in each energy bin are calculated using the Poissionian statistic.
The resulting spectrum is then grouped using the task {\it grppha}
 of the same library, so that the energy channels contain at least 20 photon
 counts.

Spectral analysis is then finally performed in the third step.
The spectrum is analyzed  with XSPEC, provided that a certain number of
constraints are satisfied. These constraints take into account the 
instrumental limits of the receiver and are described later (see Sect. 3.3).
If the spectrum does not violate any of these  constraints 
then the fit is performed
modeling the emission with a single-temperature optically thin plasma.
This emission model is the single-temperature {\it mekal} model implemented 
with XSPEC and 
has three free parameters: the gas temperature, the 
metallicity abundance and the normalization; the other parameters being kept 
fixed.
The spectrum is background subtracted using the background file 
$B_{\Delta E}^{(mk)}$ previously generated.
Errors at the $68\%$ confidence level (c.l.) for the considered parameters
 are estimated using the tasks {\it error} and {\it steppar}.

\subsection{Construction of spectral temperature data sets}
The construction of spectral samples is constrained by the
instrumental limits of the receiver.
For the considered emission models and energy bands the spectral fit procedure
described here is applied to a simulated cluster at redshift $z$ if 
a number of constraints are satisfied.
These constraints are chosen so that a mock sample of spectral temperatures
is generated which satisfies a number of properties as those of a set of 
spectral temperatures obtained from real cluster data.
These constraints are set as follows.
The signal-to-noise ratio ($S/N$) is calculated as

\be
S/N= \sqrt {t_{exp}} P  / \sqrt{ P+B},
\label{snr}
\ee
where  $P$ is the  source photon rate (photons/sec),  
$B=B_a\pi(\Delta \theta)^2$ is the photon background rate, $\Delta \theta$ 
 the source angular aperture  and $t_{exp}$ is the 
observation time. 
For global cluster temperatures the spectral samples are constructed 
using data from the simulation ensemble at redshifts $z=0.087,0.47,0.85$.
For a given energy band the exposure time is held fixed for all the 
clusters at all the redshifts, $t_{exp}=220ks$ for the energy interval
 $[0.5-10]keV$ and $t_{exp}=440ks$ for the interval $[2-10]keV$.
In order for the spectral fits to be performed, a first criterion that must
 be satisfied is that the number of source photons must exceed a
 a minimum threshold of $N_{\gamma}>250$.
Another criterion is that the cluster spectra must have 
 a signal-to-noise ratio of at least 
$S/N>10$. This threshold selects spectra of good quality, and  
at high redshifts this prevents the spectral fitting of background 
dominated spectra.

Spectral fits are performed in the energy band  $[2-10]keV$ only 
for those cluster with $T_{mw}^{200} > 2keV$.
Spectral temperature profiles are constructed from simulation data at 
redshifts  $z=0.116, 0.052, 0.039$ and $z=0.025$. For these spectra 
$t_{exp}=140ks$ for all the clusters and redshifts in the considered 
energy band $[0.5-7]keV$.
The spectrum of each 
annulus is fitted if it satisfies $N_{\gamma}>250$ and 
$B<2P$. 
The last criterion 
extracts from the ring under consideration spectra having 
$S/N$ with high values ($\simeq 10^2$).
This removes from the spectral sample fits of poor quality,
 though a large number of spectra can still be  fitted.
For global cluster temperatures the redshifts have been chosen in 
order to highlight the biases associated with the cluster selection 
owing to the detector physical limits. Therefore there are three
different and well separated redshifts rather than the redshift distribution
which would be expected  from a real cluster catalog.
For spectral temperature profiles the adopted range of redshifts 
was motivated by the need to mimic that of the clusters  for 
which temperature profiles are measured (see, for example, 
Vikhlinin et al. 2005).

The following constraints take into account the finite field of view of
the S3 chip.
For global cluster temperatures spectral analysis is performed only if 
$\theta_{\Delta}\equiv R_{\Delta}/d_A(z) < \theta_{S3}/2 =4.2 '$, 
here $d_A(z)$  is the angular distance in a flat cosmology. 
For the spectral temperature profiles a ring spectrum is fitted
only if the ring has inner radius $r_{in}/d_A(z) > 4 ''$ or outer radius
$r_{out}/d_A(z) < 4'$, and  thickness $(r_{out}-r_{in})/d_A(z) > 5 ''$.
Finally, a spectral fit temperature is rejected if, 
for the considered number of degrees of freedom (d.o.f.),
the statistic of the fitted model a gives a value of the $\chi^2$
 larger than the threshold value $\chi^2_c$ for which the null 
hypothesis probability is larger than 10 \%. 
Values of $\chi^2_c$ were pre-calculated as a function of the number
of d.o.f. up to $d.o.f.=1,000$ and stored in a table which was 
used at run time.
The range of values for the reduced $\chi^2_{\nu}\equiv \chi^2/d.o.f.$ 
of the fits lies between $0.5$ and unity, with the bulk of the 
values clustered around $0.5$.

\subsection{Construction of the Response Functions}
 The instrumental response of the ACIS-S3 detector is position dependent,
but can be considered approximately constant within each of the 32x32 square
regions covering the chip.
For extended sources, weighted ARFs and RMFs are then constructed by summing 
the corresponding subregion matrices, with a weight proportional to the 
photon count of the spectral image in the region under consideration 
\cite{ga04}.
Here a simplified treatment is adopted, in which the source emission is first
integrated spatially, and then the single source spectrum is convolved with 
appropriate template ARFs and RMFs.
For an isothermal gas the two methods give the same results, if there is 
a temperature gradient then photons far from the cluster center will 
be in proportion weighted more than in the procedure which sums over
 subregions.
Therefore, this implies on average global cluster temperatures 
biased toward lower values, with respect the first method. Because the 
bulk of the emission is concentrated at the cluster center, this effect is 
however expected to be negligible. The same arguments apply to the photon 
spectra extracted from cluster annuli, for which it is assumed that 
spatial variations of the gas temperature within a single annulus can be 
neglected.

 If one considers a large number of spectra, the weighting errors associated 
with a single convolution can be statistically reduced if the source spectra
are convolved with weighted ARFs and RMFs extracted from a template  
observation of a cluster with a regular gas distribution. 
This cluster must have the characteristic of being in a highly relaxed 
state and of covering a large field of view in the ACIS-S3 chip.
To this end Abell 2029 was chosen, for which analysis of the
ACIS-S3 image \cite{lew02} shows a very regular structure and a 
spatial extent $\simeq 4 ^{\prime}$ wide.
The cluster is located at $z=0.0767$, spectral analysis yields an
intracluster gas temperature $T_X \simeq 8keV$ and a virial radius 
$r_{vir} \simeq 2.5 Mpc$. The cluster exhibits a regular X-ray morphology and
 there is no evidence of a cooling flow at the cluster center.
A set of weighted ARFs and RMFs was generated with different
geometrical boundaries using the 
tool {\it acisspec} of the CIAO \footnote {http://cxc.harvard.edu/ciao} 
version 3.0.2, with calibration database CALDB (version 2.28). 
 The selected source regions are centered on the peak of the X-ray emission,
with their geometry being chosen to approximately fit that of the analyzed 
spectra. 
 For each source spectrum, the pair of response matrices with geometrical 
boundaries providing the best fit to those of the considered source is 
chosen in order to 
be convolved with the binned source spectrum $S_{\Delta E}(XS)$.
The same pair is later used when the spectrum is fitted 
using the single-temperature emission model {\it mekal} implemented 
with XSPEC.
The main advantage of the procedure described here is that the adopted 
approximation allows one to analyze with a reasonable computational cost a 
large number of spectra, as it is in this paper.

\section{Statistic  of cluster substructure} \label {sec:statistic}
Spectroscopic measurements of the cluster X-ray temperatures can also 
be affected by the cluster dynamical state (ME).
The amount of substructure present in the inner mass distribution
of a galaxy cluster is closely related to its dynamical state \cite{ri92}
and many statistical measures have been proposed to quantify cluster
substructure (Buote 2002 , and references cited therein).
Analysis of X-ray images does not suffer from projection effects which are
present in the optical band and the X-ray surface brightness is expected to
qualitatively follow the morphology of the projected cluster 
mass density \cite{bu02}.
For this reason, in this work is adopted the power ratio method 
\cite{bu95} as a statistical indicator of the cluster dynamical state.
The method has been widely used 
to study  cluster X-ray morphologies (cf. Buote 2000). 
An application of this method to analyze global morphologies of
simulated clusters was already performed in an earlier paper \cite{val99},
 and a systematic statistical analysis of the evolution of
cluster X-ray morphology using hydrodynamic simulations in different 
cosmological models is in preparation (Buote et al. 2006).

The method works as follows. The X-ray surface brightness 
$\Sigma_X(\rho,\varphi)$
along a given line of sight is the source term of the pseudo potential 
$\Psi(\rho,\varphi)$ which satisfies the 2-D Poisson equation. The pseudo 
potential
is expanded into plane harmonics and the $m-th$ coefficients of the expansion 
are given by :

\be
\alpha_m = \int_{R^{'}\leq R}  d ^2 x^{'}  \Sigma_X(\vec x^{'}) {R^{'}}^m
\cos(m\varphi^{'}),
\label{am}
\ee
\be
\beta_m = \int_{R^{'}\leq R}  d ^2 x^{'}  \Sigma_X(\vec x^{'}) {R^{'}}^m
\sin(m\varphi^{'}),
\label{bm}
\ee

where $\vec x^{'}=(\rho,\varphi)$ and the integration is over a circular 
aperture of radius R, which is also termed the aperture radius $R\equiv R_{ap}$.
The  $m-th$ power ratio is then defined as
\be
\Pi^{(m)} (R_{ap}) = \log_{10} (P_m/P_0)~,
\ee
where
\bea
P_m (R_{ap})& =& {1 \over 2 m^2} (\alpha_m^2 + \beta_m^2) ~~ m>0,\\
P_0 &= &[ \alpha_0 \ln(R_{ap}/{\rm kpc}) ]^2.
\eea 
The ratio $P_m/P_0$ is a measure of the amount of structure present 
on the scale of the aperture radius $R_{ap}$. For a relaxed configuration  
$\Pi^{(m)}\rightarrow - \infty$. The values of $P_m$ depend on the choice of 
the coordinate system. If the origin is the center of mass then $P_1$ 
vanishes and $\Pi^{(2)}$ is a measure of the degree of flattening \cite{bu02}. 
Large values of $P_3/P_0$ indicate asymmetric distributions and this ratio
will be used in the subsequent analysis as an indicator of the amount of 
substructure present in a cluster. The moments $P_m(R_{ap})$ of a chosen 
cluster at a given redshift are calculated
along a line of sight as in Eq. \ref{tw}, by performing the integrals 
(\ref{am}),(\ref{bm})
according to the SPH prescription. The origin of coordinates is set at the 
peak of the X-ray emission.

In order to analyze how spectral fits are affected by the amount of substructure
present in a cluster, three separate cluster sub-samples were extracted 
from the cluster sample. The sub-samples were generated by first constructing 
the cumulative distribution of the sample values of $\Pi_3$, and then
identifying  those clusters with a value of $\Pi_3$ below the threshold values
which define the $25\%$,~$50\%$ and $75\%$ percentile of the cumulative 
distribution, respectively.  
Cluster sub-samples will be 
cross-correlated with values of the spectroscopic temperatures in order to 
investigate how the cluster dynamical state can bias spectroscopic
measurements.

\section{Results and discussion} \label {sec:results}
This section is dedicated to analyzing the dependence of spectral fit 
temperatures, as obtained from simulation data using the procedures
described in sect. 3, against cluster temperatures defined according to 
weighted averages. 
Global cluster temperatures are discussed in the first two parts, projected 
temperature profiles from individual clusters are presented in the third 
part.
\subsection{Global cluster temperatures: effects of cooling}
The spectral fits temperatures $T_s^{\Delta}$ are obtained by applying the
prescriptions of sect. 3 to a cluster sample which is constructed by grouping
data from the simulation ensemble at redshifts $z=0.087,0.47$ and $z=0.85$. {Those 
clusters for which the values
of $\Pi_3$ exceed the threshold value which defines the $50\%$ percentile of the 
cumulative distribution have been removed from the cluster sample.
For those clusters which are part of a cluster sub-sample, identified by a 
given value of redshift $z$ and overdensity $\Delta$, the cluster power 
ratios were calculated according to the procedures described in sect. 4, 
 by choosing the values of the aperture radius $R_{ap}$  
to match the radius $R_{\Delta}$ ( see sect. 5.2). The power ratio 
cumulative distributions were then constructed for these cluster sub-samples.
This procedure has the effect of removing from the cluster sample  most of
those clusters undergoing a major merger event, thus  disentangling
the effects of cooling from those due to merging when analyzing the biasing
of spectral fit \tps. The choice of the threshold value is somewhat arbitrary, 
nonetheless it has been found that
the results obtained are fairly robust against changes in the chosen value.

As far as concerns the dependence of the sample on redshift, at high redshifts the 
construction of the sample is mainly limited by noise, }
for example the most massive cluster at $z=0.85$ has $r_{200}\simeq 1.2 Mpc$
and in the $[0.5-10]$keV energy band it emits $N_{\gamma} \simeq 4,200$ photons.
The photon count rate of the source is $c/s\simeq 2\cdot10^{-2} sec ^{-1}$, 
with a similar value for the background rate.
The latter depends on the source size through $B\propto \Delta \theta ^2$.
At low redshifts the sample construction is limited by the geometrical 
limits and the main contribution is given by the low 
temperature clusters. In the $[2-10]$keV bandpass the sample construction
is mainly constrained by the photon threshold $N_{\gamma} > 250$, because
of the low number of photons emitted by the clusters.
In the following analysis mass-weighted temperatures, which are supposed 
to be fair tracers of the virial values, will be taken as reference 
temperatures with which to correlate emission-weighted temperatures or 
spectral temperatures.
For a fixed cluster overdensity $\Delta$ linear fits of logarithmic variables
will be of the form 
$ \log_{10} T_{ew,s}^{\Delta} = a+b \log_{10}  T_{mw}^{\Delta}$, with errors
on spectral fit \tps given by the fit of the XSPEC {\it mekal} model
and those on the best fit parameters will be assumed at the $68 \%$ c.l..    

\begin{figure}[t]
\begin{center}
{\includegraphics*[height=12cm,width=14cm]{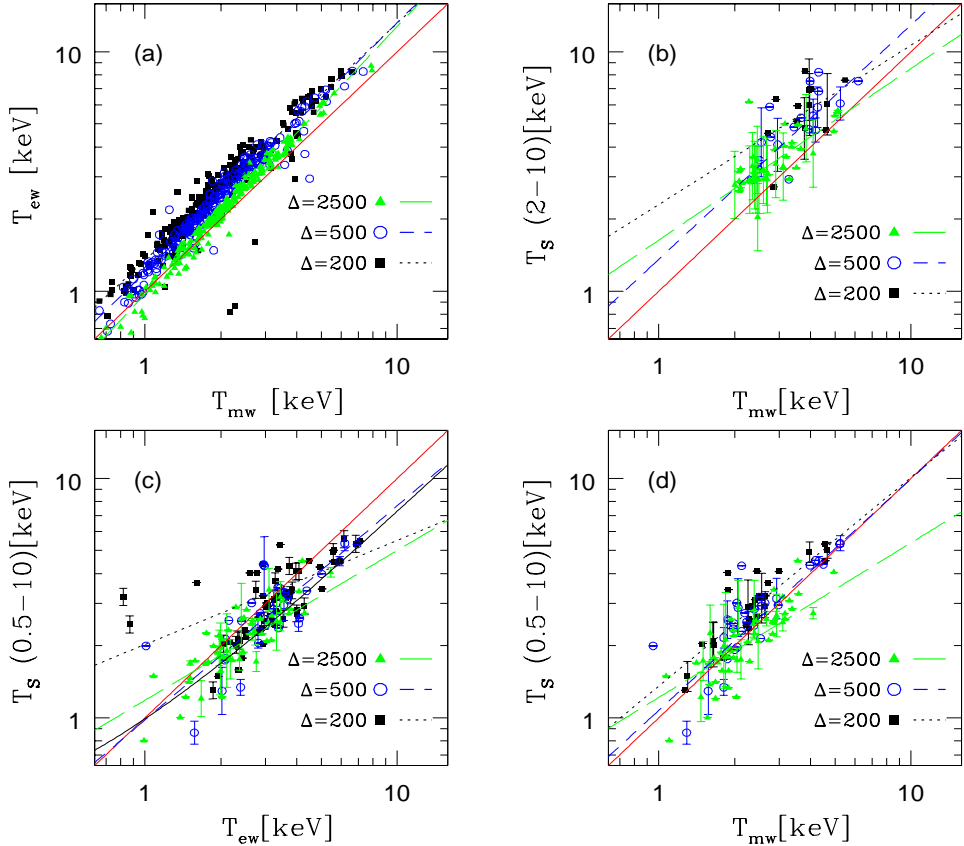}}
\end{center}
\caption{  Spectral temperatures $T_s$ are plotted 
as a function of the cluster mass-weighted ($T_{mw}$) or emission-weighted
($T_{ew}$) temperatures. Upper panel (a) refers to $T_{ew}$ versus $T_{mw}$.
For $T_s$ the numbers in parenthesis  indicate the energy band in keV.
The temperatures are found according to the procedures described in Sect. 3 
and are defined within cluster overdensities 
$\Delta=2500$ (filled triangles), $\Delta=500$ (open circles) and
$\Delta=200$ (filled squares).
The diagonal line is the line of equality, the line associated with 
the $\Delta$ symbol corresponds to the linear fit given in Table 1.
For the sake of clarity not all of the points and
not all of the error bars are shown.}
\label{fig:tws}
\end{figure}
The correlation between emission-weighted \tps and spectral \tps versus
mass-weighted \tps are displayed in Fig. \ref{fig:tws} for different
overdensities and energy bands.
Within a different panel different symbols are for different 
overdensities $\Delta$, the lines associated with the $\Delta$ symbols 
correspond to the linear fits, with the coefficients $a$ and $b$ given in 
Table 1. For the sake of clarity random subsamples are plotted.
There is a robust correlation between $T_{ew}^{\Delta}$ and  
$T_{mw}^{\Delta}$
 with  $T_{ew}^{2500}\simeq  T_{mw}^{2500}$. At smaller overdensities
this is not satisfied, with  $T_{ew}^{200}$ being systematically higher
than  $T_{mw}^{200}$. 
Emission-weighted \tps weight preferentially the inner core regions and 
change a little when going from $\Delta=2500$ to $\Delta=200$, whereas 
 $T_{mw}^{200}$ drops to smaller values when $R_{\Delta}$ gets higher.
From Table 1 $T_{ew}^{200}/T_{mw}^{200} \simeq 1.35 (T_{mw}^{200})^{-0.01}$,
so that  $T_{ew}^{200}$ is $\simeq 30\% $ higher than  $T_{mw}^{200}$, with 
a weak scale dependence on  $T_{mw}^{200}$.
These results are at variance with those of ME. A comparison with their
Fig. 5 shows that the relation  $T_{ew}$ versus $T_{mw}$ is approximately 
around the line of equality, with some preference for the scattered points
to lie above the line, i.e.  $T_{mw}\simgt  T_{ew}$. 
These differences must be ascribed to the different physical modeling  of 
the gas in the hydro simulations. In ME the gas was treated adiabatically,
 here
the runs incorporate radiative cooling and star formation, as well as SN
feedback. In the cluster central region the gas is subject to radiative
losses and cold gas which is Jeans unstable is removed because of star 
formation.  Previous analyses \cite{lew20,val02} have shown that taking
into account these effects has the consequences of changing the overall
temperature profile, with a temperature decline with the radius being steeper
than in the adiabatic runs.
This is the main reason for the discrepancy found between the behavior 
of $T_{mw}$ versus $\Delta$ here and in ME.
It must be stressed that this discussion refers to relative differences 
between mass and emission-weighted temperatures, when compared against the
corresponding  ones in the adiabatic runs.

It is useful to perform a comparison between absolute values. The relation
between mass-weighted \tps of different runs can be calibrated using 
the mass-temperature relations. For $\Delta=500$ Eq. (6) of ME gives 
$M_{500}-T_{mw}^{500}$. The analogous relation here is ( for h=1)

\be
\log_{10} E(z)M_{500}=(13.39 \pm 3 \cdot 10^{-4}) + 
(1.68 \pm 8\cdot 10^{-4} ) \log_{10} T_{mw}^{500}.
\label{mt}
\ee
To ease the notation in the following discussion the superscript
$\Delta$ is dropped from the temperatures, the letter {\it c} stands for the 
cooling runs 
performed here  and {\it a} refers to the adiabatic runs of ME. 
By equating the two mass-temperature relations one finds 

\be
T_{mw}(a)=(0.77\pm0.04) T_{mw}(c)^{1.06 \pm 0.02},
\label{ta}
\ee
this allows us to relate $T_{ew}(a)$ to $T_{mw}(c)$ using the linear fits 
coefficients of Fig. 6 of ME.  This yields
\be
T_{ew}(a)/T_{ew}(c)\simeq(0.52\pm0.03) T_{mw}(c)^{0.08\pm0.02}.
\label{trb}
\ee

So that cooling runs have emission-weighted \tps which are for $\Delta=500$ 
a factor $\simeq 2$ higher than in the corresponding adiabatic runs.
Similarly, the ratio for mass-weighted \tps is 

\be
T_{mw}(a)/T_{mw}(c)\simeq(0.77\pm0.04) T_{mw}(c)^{0.07\pm0.01}.
\label{trc}
\ee
Therefore,  mass-weighted \tps are then higher by  a factor  $\simeq 1.4$.
The ratio $T_{ew}/T_{mw}$ is higher than in ME by a factor $\simeq 
30 \% $, with a weak dependence on $T_{mw}$. It is found
\be
\left [ \frac {T_{ew}} {T_{mw}} \right ]_{(a)} /\! 
\left [ \frac {T_{ew}} {T_{mw}} \right ]_{(c)} 
\simeq 0.67 T_{mw}(c)^{0.125 \pm 0.01}.
\label{trd}
\ee
These relations show how incorporating into the simulations a more 
realistic physical modeling of the gas leads to differences in the gas
temperature and distribution which can have a strong impact on temperature
averages.
This suggests that also the relationships between spectral \tps and mean 
gas \tps can be modified when the simulations take into account
 the effects of radiative cooling and star formation.
The relationships between spectral fit \tps and averaged gas \tps are
 shown in the other panels of Fig. \ref{fig:tws}.

\begin{figure}[t]
\begin{center}
{\includegraphics*[height=8cm,width=14cm]{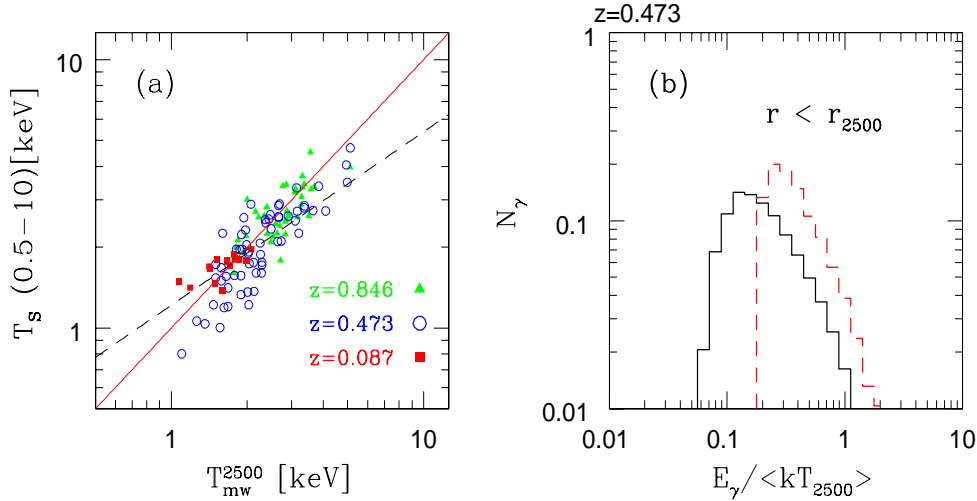}}
\end{center}
\caption{In panel {\bf (a)} it is shown for $\Delta=2500$ the relation
between $T_s[0.5-10]$ and $T_{mw}^{2500}$ of Fig. \ref{fig:tws}(d), different
symbols refer to clusters selected at different redshifts. The dashed line 
is the linear fit of Table 1. {\bf (b)}: The source photon distribution emitted
within $R_{2500}$ at $z=0.473$ is binned as a function of the photon energy
in units of the average gas temperature. Continuous (dashed)
 line is the average over the 20 most (least) massive clusters of the sample.
Vertical units are arbitrary.}
\label{fig:his}
\end{figure}

\begin{table}[t]
\label{tc}
\begin{center}
\begin{tiny}
\caption{ Values of the best-fit coefficients and $1\sigma$ confidence limits
 for linear fits of the form $\log T_w^{\Delta}=a+b \log T_w^{\Delta}$ 
applied to the cluster emission-weighted, mass-weighted and spectral 
temperatures in keV units. 
The coefficients are calculated for cluster temperatures 
defined within radii enclosing cluster overdensities 
$\Delta=2500,~500,~ 200$. The fits are performed over samples which are
constructed grouping data from the simulation ensemble at redshifts 
$z=0.087,0.47,0.85$.
The subscript $w=ew,mw,s$ indicates the weighting schemes, described in 
Sect. 3. For spectral temperatures the numbers in square  
brackets indicate the energy band. An asterisk in the column of the linear
 fit coefficient $a$ means that the $\chi^2$ probability  of the 
fit is below $0.01$.}
\begin{tabular}{cccccccccccc|l}
\hline \hline
\multicolumn{4}{c}{$\Delta=2500$}
&\multicolumn{4}{c}{$\Delta=500$}
&\multicolumn{4}{c}{$\Delta=200$}
&\multicolumn{1}{l}{Temperature }\\
& & & &
& & & &
& & & & {relations}\\
\hline
a&$\sigma_{a}\cdot10^2$&b&$\sigma_b\cdot10^2$&
a&$\sigma_{a}\cdot10^2$&b&$\sigma_b\cdot10^2$&
a&$\sigma_{a}\cdot10^2$&b&$\sigma_b\cdot10^2$& \\
\hline
-0.013&0.4&1.12&1&
0.08&0.5&1.05&1.5&
0.13&0.8&0.99&2.3& $T_{ew}$-$T_{mw}$\\

 0.07 & 0.8 & 0.63 & 2.6 &
-0.009 & 3.5& 0.9& 5.7&
 0.3$(*)$ & 3.5 & 0.43 & 6 & $T_{s[0.5-10]}$-$T_{ew}$\\

 0.085$(*)$ & 0.8& 0.64&2.8 &
 0.03 & 3& 0.97& 5.8& 
0.13 & 4 & 0.87 & 8 &$T_{s[0.5-10]}$-$T_{mw}$\\

 0.21 & 5.2& 0.72& 13&
0.14  & 23 & 0.98 & 37 &
0.36  &35 & 0.66  & 58 &$T_{s[2-10]}$-$T_{mw}$\\

\hline
\end{tabular}
\end{tiny}
\end{center}
\end{table}
From Fig. \ref{fig:tws}(d) it can be seen that in the energy band 
$[0.5-10]$ keV
the relationship between $T_s$ and $T_{mw}$ for $\Delta=2500$  
 has many points which are approximately distributed 
according to the linear regression fit, 
nonetheless the $\chi^2$ probability of the fit is below $0.01$.
This follows because, as a consequence of selection effects, 
there are many points of good statistical 
quality which are at low temperatures.
The best-fit parameters are mostly weighted by these points, 
and the best-fit line follows their distribution.
This is more clearly illustrated in Fig. \ref{fig:his}, where in the left 
panel (a) are shown all of the points with which the 
spectral sample
$T_s[0.5-10]-T_{mw}$ for $\Delta=2500$ is constructed. 
Only spectral temperatures with  $\chi^2$ values from the spectral fits
below the threshold value $\chi^2_c$ have been plotted.
For a given number of d.o.f. $\chi^2_c$ is such that
the $\chi^2$ probability of the fit  gives $P_{\chi_c^2}(d.o.f.)=10\%$.
As previously outlined at high 
z's the construction of the sample is constrained by noise, with 
only massive clusters being selected, whereas at low redshifts cool
clusters are preferentially selected.  
The poor quality of the linear fit is mainly determined by the relatively
 large number of outliers which are present at high z's 
in the spectral sample and which do not follow the distribution of 
the low temperature clusters. This feature is closely related to the way 
in which the introduction of cooling affects the gas distribution. 

ME argue that spectroscopic \tps $T_s[0.5-10]$ are biased toward lower values
of the mass-weighted \tps because spectroscopically determined \tps 
are weighted by the fitting process according to the photon counts, 
which are dominated by the low-energy part of the spectrum.
From the spectral sample of Fig. \ref{fig:his}(a) it can be seen that for massive
clusters spectral \tps are lower than mass-weighted temperatures, whereas 
$T_s\simeq T_{mw}$ as cool clusters are considered.
This behavior follows from two effects: the way in which the sample is 
constructed and the introduction of the physical modeling of cooling in  the
simulations. 
At high redshifts the sample is dominated by massive clusters for which 
$T_s\simlt T_{mw}$, as in ME, at low redshifts the sample population is 
dominated 
by cool clusters. For these clusters, according to the cooling scenario
 \cite{br00,vo01}, the efficiency of galaxy formation is higher than 
in hot clusters. This implies a removal of the low-entropy cooled gas, 
transformed into stars, and a subsequent inflow of the surrounding high-entropy
gas. Therefore, in this scenario, it follows that for cool clusters the
central cluster temperature, in units of a characteristic cluster temperature, 
is higher than for massive clusters.
This issue will be discussed in more detail in the section dedicated to 
analyzing the temperature profiles. 
Fig \ref{fig:his}(b) shows the energy distribution of the photons 
emitted at $z=0.473$ within $R_{2500}$ by the gas particles   
 of the the 20 most (least) massive clusters of the spectral sample.
Energy is in units of the averaged gas temperature within $R_{2500}$ :
$<T_{2500}>= \sum_i T_i/N$, where the summation is over all of the N gas 
particles within the radius.

The histograms of the two distributions clearly indicate that cool
clusters will have average temperatures higher, when rescaled to a 
characteristic mean gas cluster temperature, than the corresponding ones
of massive clusters.
Spectroscopic temperatures too are expected to follow this behavior and, 
as a consequence, will have a dependence on mass-weighted \tps different from 
that found by ME. The argument can be quantified in more detail by 
applying again the same arguments which lead to Eq. \ref{ta}. From Fig. 6 
of ME it is found in the $[0.5-10]$ bandpass for $\Delta=500$  
\be
T_s(a)=(0.6\pm0.05) T_{mw}(c) ^{(1.1\pm0.03)},
\label{tre}
\ee
and

\be
T_s(a)/T_s(c)\simeq(0.56\pm0.05) T_{mw}(c) ^{(0.13\pm0.06)}.
\label{trf}
\ee
Using this equation $T_s(c)\simeq1.8T_s(a)$ for $T_{mw}=1$keV but 
$T_s(c)\simeq 1.3 T_s(a)$ at  $T_{mw}=10$keV.
This shows that incorporating cooling into the simulations has the net
effect of introducing for spectral \tps a scale dependency much stronger 
than in the adiabatic runs of ME.
The runs performed here take into account the  dependence of the cooling 
function on the gas metallicity; line emission from cold gas is expected
to bias significantly spectral \tps toward lower values (ME), from the results
found here it turns out that this bias is largely covered by the scale 
dependency introduced by cooling.
The scale dependency of the $T_s-T_{mw}$ relationship is also dependent
on the energy bandpass, as it is found in ME, when passing from the 
energy interval $[0.5-10]$keV to $[2-10]$keV panel (b) of Fig. \ref{fig:tws}
shows that cool clusters have now $T_s\simgt T_{mw}$ at $\Delta=2500$.
This follows because the low energy photons which weight spectral
\tps towards lower values are removed when the spectral fitting is 
performed in the  $[2-10]$keV range.

These combined effects on the scale dependency of $T_s$ versus $T_{mw}$ 
are equally present when the cluster radius $R_{\Delta}$, within which the
\tps are found, is increased. Here the scale dependency is also affected 
by the biases associated with the cluster selections because of the 
geometrical constraints that limit the sample construction.
Thus, the cluster sample with $\Delta=200$ of Fig. \ref{fig:tws}(d) is expected 
to be
less populated by massive clusters than the corresponding sample with 
$\Delta=2500$. This is because most of the clusters at $z=0.087$ have
$\theta_{200}$ greater than the field of view of the S3 chip.
This bias is relatively unimportant in the determination of the slope
of the  $T_s-T_{mw}$ relationship for $\Delta=200$, which is dominated 
by the general tendency of having spectral \tps higher than mass-weighted 
temperatures. This behavior was already detected when discussing the relationship 
between $T_{ew}$ and $T_{mw}$ versus $\Delta$ and it is a consequence
of incorporating cooling into the simulations.
According to the simulation results mass-weighted \tps have a strong
dependency on $\Delta$, whereas $T_s$ or $T_{ew}$ remain relatively unaffected
when $R_{\Delta}$ is increased because their weighting scheme favors the 
central regions where the bulk of the emission is located.
Therefore it turns out that spectral \tps are systematically higher 
than mass-weighted \tps when $\Delta\simlt 10^3$.
These results are at variance with those of ME, for which spectral fit 
\tps are biased against mass-weighted \tps by a $\simeq 20\%$ 
toward lower values.

 There is also a well defined tendency to have large statistical errors
 on the spectral fit \tps when the overdensity $\Delta$ decreases. 
This effect is present in all of the panels of Fig. \ref{fig:tws}, where 
a spectral $T_s$ is correlated with a weighted average gas temperature.
The errors $\sigma_a$ and $\sigma_b$  on the best fit parameters 
are given in Table 1 and indicate this tendency.
These large errors on the  spectral $T_s$ as $R_{\Delta}$
is increased are likely to be induced by the inclusion within the spectral
fit volume of an increasing amount of gas at \tps lower than those
at the cluster center.
In the $[2-10]$keV bandpass the cluster spectral \tps with $\Delta=200$ 
(Fig. \ref{fig:tws}(b)) have large statistical errors. The slope of the 
$T_s-T_{mw}$ relationship is 
greater than one, but with a statistical error as large as the slope itself.
As a general rule it is found that in the energy range $[2-10]$keV, 
spectral fits \tps have large statistical errors. This is mainly due to the
poor quality of the photon statistics associated with the fits. 
This is closely related to the construction of the spectral temperature data
 sets. There is an important difference between the way in which spectral
temperature data sets are constructed here and in ME. 
Each spectrum is fitted by ME, using an isothermal spectral fit model {\it 
mekal} with fixed metallicity, keeping the number of observed photons 
constant to $20,000$. 
At variance with ME, here the exposure time is kept fixed, so that 
the observed number of photons can vary from cluster to cluster.
This choice was motivated by the request of creating spectral temperature
sets from fake observations with realistic exposure times, therefore 
avoiding the selection biases associated with the inclusions in the 
spectra data set of cluster spectra with an otherwise low photon count.
As a consequence, spectral fits in the  $[2-10]$keV energy band have 
generically a poor statistic, because of the smaller number of photons with
 respect to the  $[0.5-10]$keV band, even with the very high exposure time 
(see sect. 3.3) of $440$ks.

In Fig. \ref{fig:tws}(c) for different overdensities $\Delta$ the 
relationships between $T_s[0.5-10]$ and $T_{ew}$ are displayed. 
These have the same 
scale dependencies as those between $T_s$ and $T_{mw}$.
The correlations between $T_s[0.5-10]$ and $T_{ew}$ can then be explained using
the arguments previously discussed. As a general result it appears that 
$T_s$-$T_{ew}$ relationships are more widely scattered around the line 
of equality than their counterparts  $T_s$-$T_{mw}$.
This effect is particularly severe when the overdensity $\Delta=200$ is considered.
This is most likely due to the combined effects of the 
 inclusion within the spectral fit volume of clumps of cold dense gas as 
$R_{\Delta}$ increases and to the $\rho^2$ dependence of the $T_{ew}$ measure.
 There is a tendency
for the ratio of $T_s$ to $T_{ew}$ to be higher than that of  
$T_s$ to $T_{mw}$, because of the biasing between $T_{ew}$ and $T_{mw}$.
A comparison with previous results can be performed for the relation
$T_s[0.5-10]$-$T_{ew}$ at $\Delta=500$. From a cluster sample obtained from a 
set of hydrodynamical simulations, with a physical treatment of the gas 
similar to that used here, Rasia et al. (2005) found a relationship
at $\Delta=500$ between $T_{ew}$ and a suitably averaged temperature $T_{sl}$.
The latter is supposed to be a good approximation to the spectroscopic
temperature $T_s[0.5-10]$. A linear fit of the form 
$T_{sl}=aT_{ew}+b$ yields $a=0.7\pm0.01$ and $b=0.29\pm0.05$. 
This relationship is in good agreement with the corresponding one in Table 1.
At $T_{ew}=1keV$ the values of  $T_s[0.5-10]$ and $T_{sl}$
 are nearly equal,  whereas at $T_{ew}=10keV$ there is a $\sim 10 \%$ difference
between the two values.

\begin{figure}[t]
\begin{center}
{\includegraphics*[height=6cm,width=14cm]{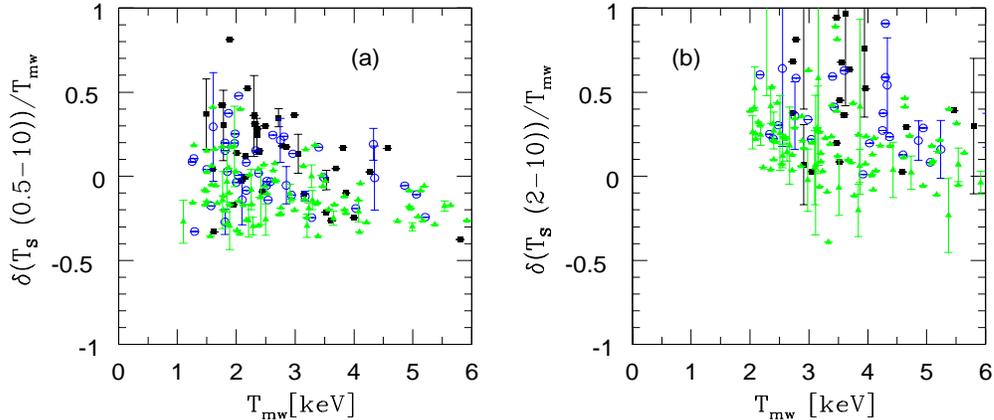}}
\end{center}
\caption{  Relative differences between spectral or emission-weighted 
temperatures versus mass-weighted temperatures. {The temperatures are found as in 
Fig. \ref{fig:tws} and 
different symbols correspond to different cluster overdensities :
$\Delta=2500$ (filled triangles), $\Delta=500$ (open circles) and
$\Delta=200$ (filled squares). The notation $\delta(y)/x$ stands for 
$(y-x)/x$.}  As in Fig. \ref{fig:tws}, random subsamples are 
plotted to avoid overcrowding.}
\label{fig:dtw}
\end{figure}

 The scale dependencies of the quantities shown in Fig. \ref{fig:tws} are more
clearly illustrated in Fig. \ref{fig:dtw}, where  
the fractional differences between correlated quantities are plotted.  
Panel (a) shows that the quantity $\delta(T_S(0.5-10))/T_{mw}$ for 
$\Delta=2500$ is characterized by negative values for $T_{mw} \simgt 3keV$, 
and by a plume
of positive values as long as  $T_{mw} \rightarrow 1keV$. This feature is 
exacerbated  when considering spectral \tps in the energy band $[2-10]keV$ 
( panel (b) ), as expected because of the lack of low-energy photons.
These results can be compared with Fig. 8 of ME, for which the condition 
$T_S <T_{mw}$ is always valid in the $[0.5-10]keV$ bandpass.
The behavior of $\delta(T_S(0.5-10)/T_{mw}$ for $\Delta=2500$ is reproduced 
also when considering \tps  defined within overdensities $\Delta=500$ or 
$\Delta=200$, here the distributions are characterized by a smaller number of
massive clusters because of selection constraints.
It is clear that the scale dependency of spectral \tps against mass-weighted 
\tps is dominated by the effect of introducing cooling in the simulations.
The biasing of spectral \tps associated  with line emission from cool gas 
was dominant in previous runs which considered only gas shock-heating, 
but it appears now  as a secondary effect.

\subsection{Global cluster temperatures: effects of substructure}
The discussion of the relationship between spectral and 
mass averaged \tps has not yet considered the role of merger events  
during the individual cluster formation history.
According to ME accretion and merger events will alter the 
measurement of spectral temperatures. These are expected to be significantly
biased during a major merger event. As the gas gets shock-heated 
because of the collision its temperature and luminosity will 
temporarily increase.  The merging will proceed with the subclump of 
cold gas moving toward the cluster center and with its presence will 
bias the cluster spectrum toward lower temperatures. The gas of the clump
is expected to be cooler than most of the gas in the cluster because 
of its smaller mass with respect to that of the cluster. ME argue that 
spectral \tps will be much smaller than mass-weighted \tps in the 
proximity of a merger event.

\begin{figure}[t]
\begin{center}
{\includegraphics*[height=6cm,width=14cm]{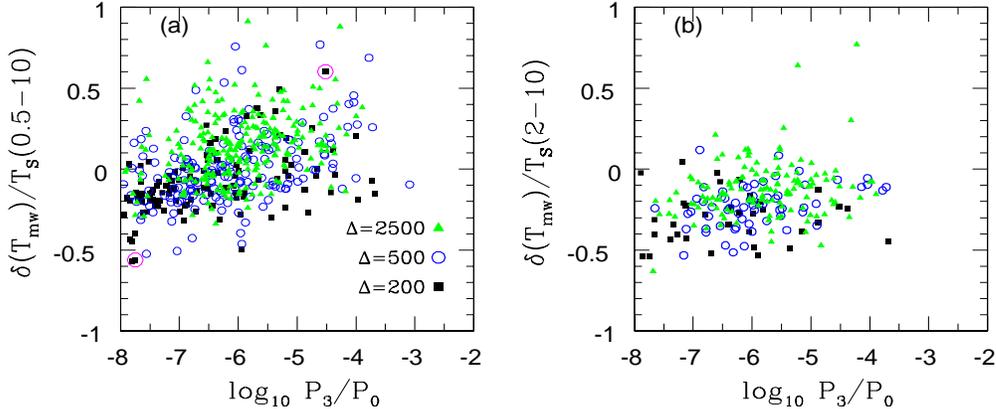}}
\end{center}
\caption{  
Relative differences between mass-weighted temperatures and spectral
temperatures are shown versus the cluster power ratios $\Pi_3=log_{10}
P_3/P_0$. {The temperatures are defined as in 
Fig.  \ref{fig:tws}
within different cluster overdensities $\Delta$.}
The values of $\Pi_3$ are calculated at the same redshift
and line of sight of the corresponding cluster temperatures.
The aperture radii are defined according to the value of $\Delta$,
with $R_{ap}=R_{200}$, $R_{ap}=R_{200}/2$ and $R_{ap}=R_{200}/4$
chosen to approximately match the value of $R_{\Delta}$ 
in correspondence of $\Delta=200,500,~2500$, respectively.
The two clusters identified by a circle in panel (a)  have their 
phase space diagram plotted in Fig. \ref{fig:plph}.}
\label{fig:dtpw}
\end{figure}

The substructure statistic used here is the power ratio method described 
in sect. 4. This statistic is different from the one used by ME  and has 
the main advantage of giving a measurement of the amount of structure present 
on a given scale, chosen accordingly to the value of the aperture radius 
$R_{ap}$. This allows us to correlate, for different overdensities or radii 
$R_{\Delta}$, the fractional differences $\delta(T_{mw})/T_S$ as a 
function of the power ratio $\Pi_3=\log_{10} P_3/P_0$. 
For the clusters under consideration, the values of $R_{ap}$ have been chosen 
to match that of the radius $R_{\Delta}$.
The two panels of Fig. \ref{fig:dtpw} refer to  $\delta(T_{mw})/T_S$ as a 
function of $\Pi_3$ for $T_S$ in the energy bands $[0.5-10]$ and $[2-10]keV$.
In each panel the distributions as obtained by considering the
three different overdensities previously considered are displayed.
A striking result of Fig. \ref{fig:dtpw}(a) is the robust correlation which 
is found between $\delta(T_{mw})/T_S$ and the substructure as measured by
$\Pi_3$. A result which is likely due to the theoretical framework of the 
statistical method used to measure substructure, which has the advantage of 
introducing well defined scale dependent quantities.
 In the $[2-10]keV$ bandpass there is a significant correlation only for 
 $\Delta=2500$ and $\Delta=500$ , the other distribution ($\Delta=200$) 
being scattered similarly to the corresponding one of Fig. \ref{fig:dtw}(b).

\begin{figure}[t]
\begin{center}
{\includegraphics*[height=8cm,width=16cm]{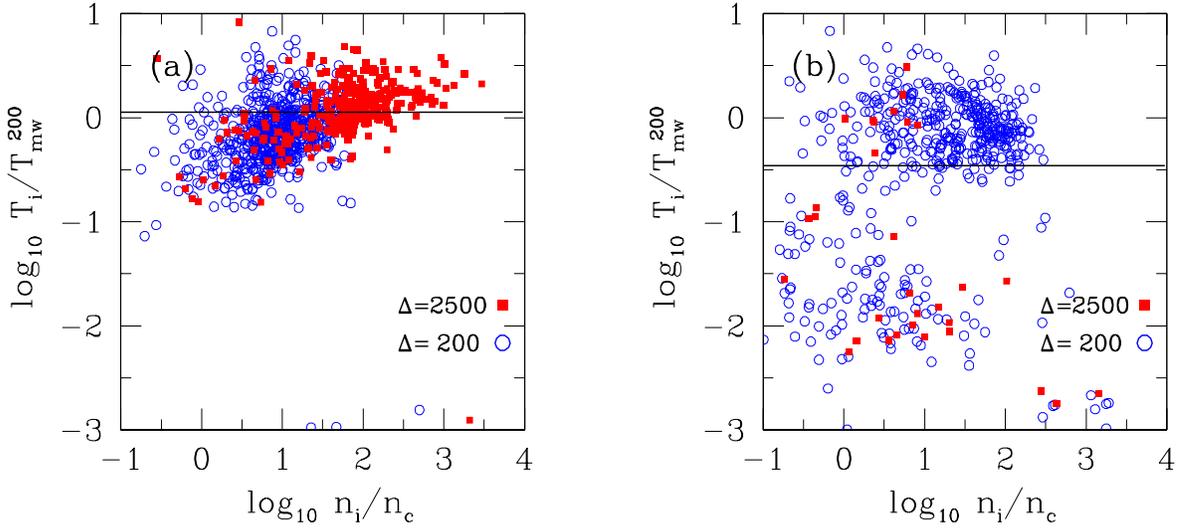}}
\end{center}
\caption{  
Phase-space diagrams of the two clusters identified in Fig. \ref{fig:dtpw}(a)
by the circles drawn around the plotted points. Left panel (a) is for the
cluster lying at the bottom left part of the panel. Temperatures are in units
 of $T_{mw}^{200}=1.8~(5.8) keV$ for the left (right) 
cluster. The number density ratios are defined as 
 $n_i/n_c=\rho_i/\rho_{c}(z)$. Filled symbols are for gas particles within
$R_{2500}$ and open symbols for those within $R_{200}$. The continuous line 
gives $T_i=2keV$.}
\label{fig:plph}
\end{figure}

The correlations are  confirmed from the values of Table 2 in which are 
reported the Spearman rank 
correlations and significance levels of the corresponding panels.
 The correlation behavior reveals that $T_S$ is significantly lower than 
mass-weighted \tps as $\Pi_3$ approaches zero, i.e. whenever the cluster gas 
distribution  is strongly perturbed, which is an indicator of a merging event.
As long as $\Pi_3$ becomes very large and negative the cluster is in a
relaxed state and spectral \tps $T_S$ are higher than mass-weighted 
temperatures.
For  $\Delta=200$ two clusters have been chosen as representative 
of these two regimes, they are identified by the circle drawn around their
points; their  phase-space diagrams are illustrated in Fig. \ref{fig:plph}.

The phase-space diagrams of the two clusters in a different dynamical state
are displayed in Fig. \ref{fig:plph}. Gas particle temperatures $T_i$ are in units
 of $T_{mw}^{200}$ and the solid line gives $T_i=2keV$. 
Left panel (a) is for the cluster 
identified by the circle lying at the bottom left corner of Fig. \ref{fig:dtpw} (a).
The cluster is in a fairly relaxed state with $T_{mw}^{200}=1.8keV$, 
$\Pi_3=-7.75$ and will be termed as 'quiescent'. In the right panel (b) 
the $\log T- \log n$ points for the cluster marked by the circle in the top
 right part of Fig. \ref{fig:dtpw}(a) are plotted. For this cluster $T_{mw}^{200}=5.8keV$  and
$\Pi_3=-4.52$. The cluster can be considered strongly asymmetric along the 
chosen line of sight, its value of $\Pi_3$ being above the threshold value
 defining 
the $75 \%$ percentile of the cumulative distribution.  
This cluster will be denoted as 'active'.
The main difference between the two distributions is the long tail of 
cool gas which characterizes the active cluster, for which $T_s\simeq 0.6 
T_{mw}^{200}$. Note that the quiescent cluster has most of the material at 
\tps below the solid line indicating $2keV$, nonetheless for this cluster 
 $T_s\simeq 2 T_{mw}^{200}$. 

These findings support the analysis of ME, for whom clusters with an 
undergoing merging activity have spectral \tps smaller than mass-weighted
temperatures. It must be stressed that the scale dependencies of spectral 
\tps which are shown in the panels of Fig. \ref{fig:dtpw} are independent 
 from those function of mass-weighted \tps which follow by introducing 
cooling in the simulations. 
\begin{table}[t]
\label{dt}
\begin{center}
\caption{ Spearman linear correlation coefficients $r_s$ and significance
levels $P_{rs}$ are given for the fractional temperature differences 
displayed in the two panels of Fig. \ref{fig:dtpw} as a function of the 
cluster power ratios $\Pi_3$.
Values of $P_{rs}$ below $10^{-2}$ have been rounded to zero and
a significance level of $5\%$ is used to reject the null hypothesis
 $r_s=0$.}
\begin{tabular}{cccccc|l}
\hline \hline
\multicolumn{2}{c}{$\Delta=2500$}
&\multicolumn{2}{c}{$\Delta=500$}
&\multicolumn{2}{c}{$\Delta=200$}
&\multicolumn{1}{l}{fractional temperature - power ratio relations }\\
\hline
$P_{rs}$&$rs$&
$P_{rs}$&$rs$&
$P_{rs}$&$rs$&\\
\hline
0.00&0.31&
0.00&0.38&
0.00&0.47& $\delta (T_{mw})/T_s[0.5-10]- log_{10} P_3/P_0$\\

0.01&0.21&
0.00&0.39&
0.65&0.08& $\delta (T_{mw})/T_s[2-10]- log_{10} P_3/P_0$\\

\hline
\end{tabular}
\end{center}
\end{table}
According to these results, the biasing of spectral \tps can be described in
terms of a two-parameter model. A first scale dependency is introduced 
by cooling and is a function of the cluster mass or mass-weighted 
temperature.
The second dependency is correlated with the amount of cluster substructure
and is independent from the first. Given the importance of this argument it 
has been decided to look at the phase-space diagram of a cluster with its 
position in Fig. \ref{fig:dtpw}(a) as close as possible to the one identified by 
the active cluster, but characterized by a low value  of $T_{mw}^{200}$.
 A cluster was chosen with $T_s\simeq 0.7 T_{mw}^{200}(\simeq 1keV)$ 
and $\Pi_3 = -4.36$. For this cluster, a distribution of 
points in the $\log T- \log n$ plane very similar to that of the active 
cluster was found.
This confirms that the two processes which govern the biasing of spectral 
temperatures, subclump accretion and radiative cooling, introduce scale
dependencies which can be considered independent of each other.

Finally, there is a further source of biasing which is due to selection 
constraints because of the detector configuration. At high redshifts the 
sample construction is limited by noise, whereas at low redshifts the 
detector geometry is dominant.
The relationships between spectral and mass-weighted \tps are then shaped 
by these selection effects, which weight the scale dependency introduced 
by cooling according to the cluster redshift.
Spectroscopic measurements with other X-ray telescopes will give 
relationships
different from those obtained here using the {\it Chandra} ACIS-S3 detector, 
though qualitatively similar. Because of the limits to the instrument 
sensitivity the generic population of a cluster sample will be dominated 
at low redshifts by cool clusters and at high redshifts by massive clusters.
Another issue concerns the geometrical limits of the detector, if its field 
of 
view is larger than that of the chip S3 spectral relationships for 
$\Delta=200$ are expected to include a  number of massive clusters 
larger than the one of Fig. \ref{fig:tws} (d). 
This in turn implies a scale effect which will change the bias between 
$T_S$ and $T_{mw}$. This effect is however expected to be negligible
since most of the new clusters will be characterized by $T_s\simeq T_{mw}$.
Thus the scale dependencies of spectral \tps found here are expected 
to be qualitatively reproduced with other detectors.

\subsection{Radial temperature profiles}
In this section the behavior of the radial temperature 
profiles as obtained from the sets of simulated spectral samples
constructed according to the procedures described in Sect. 3 are 
discussed. 
Spectral temperature profiles $T_S(r)$ are compared against projected 
emission-weighted temperature profiles $T_{ew}(r)$, the spectral fits
are performed in the energy band $[0.5-7]keV$ and the \tew \tps are 
defined in the same energy band. The \tew profiles have been calculated 
keeping fixed the chosen line of sight and the radial binning, in units of 
$r_{200}$, for all of the clusters. For a given cluster, the corresponding
 spectral  temperature profile is obtained by fitting the photon spectra of
 the considered rings. Because of the scale dependencies previously 
discussed, the sample of temperature profiles has been subdivided by
grouping individual profiles into sub-sample according to several 
cluster properties. A cluster is part of a subsample denoted by 'hot' 
if its value of $T_{mw}^{200}$ exceeds $4keV$, 'normal' if  
 $T_{mw}^{200}> 2keV$ or 'cool' when $T_{mw}^{200}<2keV$.
Moreover, the subsamples are also separated according to the degree 
of regularity of the gas distribution of its cluster members. 
A cluster is part of a subsample denominated 'quiescent'  if the value 
of $\Pi_3$ is below the threshold value which defines the $25\%$ 
percentile of the cumulative distribution of the power ratios. 
Similarly, clusters which are members of the 'active' subsample have their
value of $\Pi_3$ above the threshold which defines the $75\%$ of the 
percentile of the distribution.

For a given subsample, the mean spectral temperature profile is defined 
by averaging the profiles of all the clusters of the subsample, with the 
constraint that a profile is part of the mean if there are at least 
five spectral \tps with contiguous radial bins which satisfy the 
constraints of Sect. 3.3. This criterion was  introduced because it 
has been found that otherwise there would have been clusters with very few
annulus bins which would have taken part in the construction of the mean 
of the profile. These clusters with a very sparse sampling of their 
profile lead to a distortion of the average profile owing to the scale 
dependency introduced by cooling.
 If a cluster spectral profile is accepted to be part of the mean 
of a given subsample, then the corresponding cluster \tew temperature
 profile is also part of the average \tew profile for the subsample 
under consideration.
Averages have been performed by rescaling the cluster \tps to 
\tmw or $T_{ew}^{200}[0.5-10]\equiv T_X$, so that it is the rescaled 
profile which is averaged. This allows us to compare consistently 
mean spectral profiles with mean \tew profiles.
 In order to properly compare with available data profiles, the cluster 
temperature $T_X$ has been calculated by excluding a central region of 
size $50 h^{-1}kpc$.

The upper panels of Fig. 6 show the projected spectral temperature
profiles as a function of $r/r_{200}$. Within a given panel points with 
different symbols refer to mean spectral profiles as obtained from 
different subsamples previously defined; dashed lines are the mean \tew 
profiles extracted from the  same subsample.
The profiles of quiescent clusters are displayed in the right panel, those 
of the active cluster in the left panel.
The top right panel of Fig. 6 shows that quiescent clusters have scaled
 profiles which rise toward the cluster center, reaching their peak values 
at $r \simeq 0.02 r_{200}$ and with a steep decline thereafter.
Spectral profiles follow this behavior, but with a biasing which is 
strongly dependent on the chosen subsample. For hot clusters 
$T_S(r) \simeq T_{ew}(r)$ whereas for cool clusters this is valid only in 
the inner regions. Moving outward from the peak value the bias can be as 
high as $\simeq 20 \%$. The discussions of the previous sections suggest
that these biasing dependencies can be explained as follows: spectroscopic
measurements are biased by line emission toward lower values than 
\tew temperatures, this effect is negligible for hot clusters but 
relevant for those clusters of the $T<2$ subsample. 
This biasing is strongly suppressed at small radii because of the cooling 
efficiency for these clusters, which has removed most of the cold gas 
at the cluster cores. The same arguments apply to the active clusters 
of the left panel. Here, however, the profiles are shallower than those
 of the right panel and the peak heights are much more modest. Normal and 
hot clusters have their profiles approximately isothermal,
in the cluster inner regions spectral temperature profiles are now also 
biased with respect emission-weighted profiles. 
These dependencies of the shape of the profiles on the value of $\Pi_3$
indicate that the effects of merging on the gas distribution of the 
clusters are the main source for the differences in the profiles.
According to this framework  active clusters have profiles much 
shallower than quiescent clusters because their cores have accreted 
from subclumps a significant amount of cool gas through a number of merging
events.
For small values of the aperture radius ($R_{ap}\simlt r_{200}/4$) the 
differences between the shape of the profiles from active and quiescent 
clusters
are not as well defined as those displayed in Fig. 6.
This suggests that cools gas can significantly accrete into cluster cores 
only through major merging events, where the mass of the subclump is 
 a significant fraction of the cluster mass.

\begin{figure}[t]
\begin{center}
{\includegraphics[height=14cm,width=16cm]{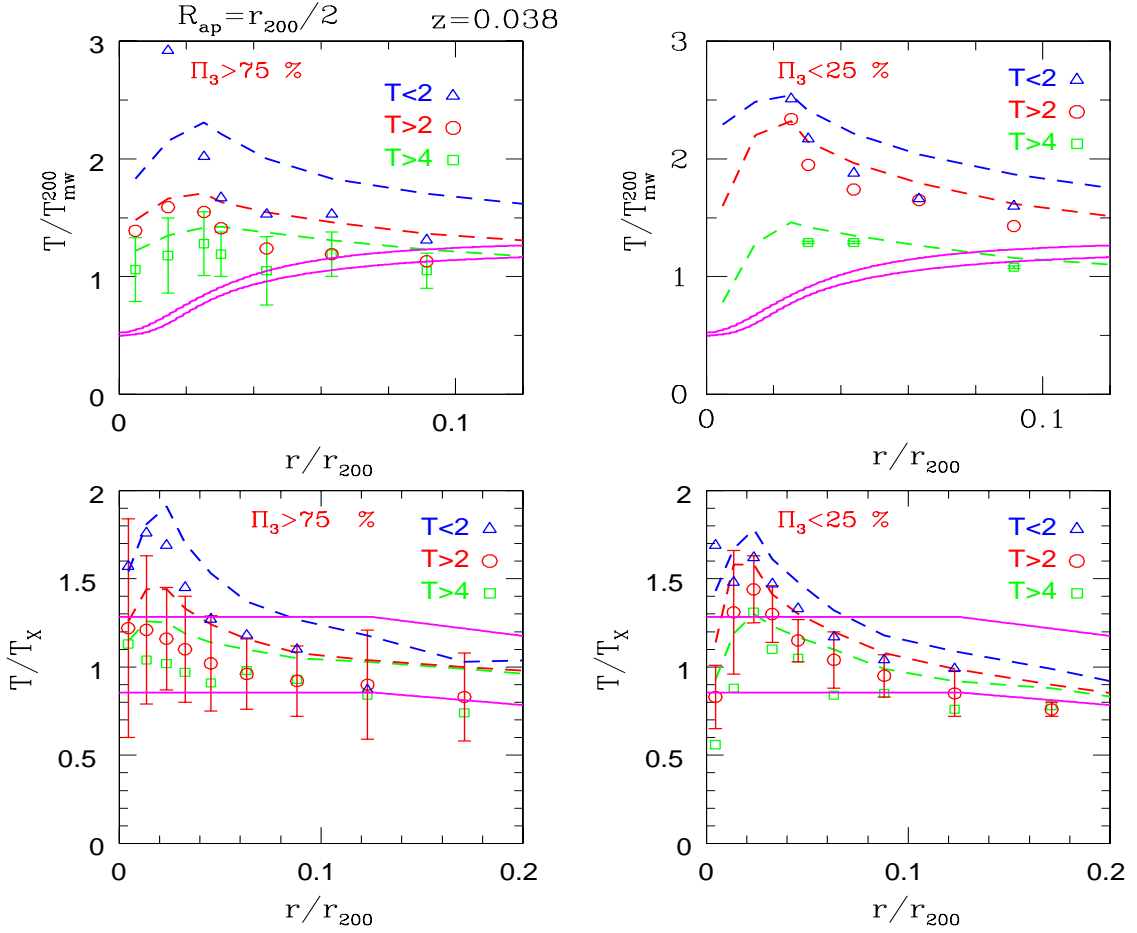}}
\end{center}
\begin{tiny}
\vspace{-1cm}
\caption{  
Spectroscopic projected temperature profiles (points) are compared
against \tew profiles (dashed  lines) and best-fit data constraints
(solid lines). The dashed lines follow from bottom to top the same subsample
order of the spectral points. The  profiles are the averages over individual 
profiles of those clusters which are part of a subsample. These 
have been defined as follows : the notation $T>4$ means that 
all the clusters with $T_{mw}^{200}>4keV$ are part of the subsample, 
while $\Pi_3 > 75\%$ means that these clusters have a value of 
$\Pi_3$ above the
threshold which defines the $75\%$ of the cumulative distribution.
The meaning of the other notations is similar. 
The power ratios have been calculated with $R_{ap}=r_{200}/2$.
The upper panels have 
profiles extracted from the spectral sample of clusters at redshift 
$z=0.038$. The profiles of the lower panels have been averaged over 
samples at redshifts $z=0.116, 0.052, 0.039$ and $z=0.025$. 
The solid lines of the upper panels define the $68\%$ c.l. of 
the rescaled best-fit profile $T(r)/{T_{mw}^{2500}}$ of Allen et al. 
(2001a). For the lower panels the solid lines give the $68\%$ uncertainties
of the best-fit profile to the data of Fig. 18 Vikhlinin et al. (2005); 
temperatures are in units of $T_X$, the cluster \tew temperature in the 
energy band $[0.5-10]keV$.}
\end{tiny}
\label{fig:tprofspe}
\end{figure}
The solid lines in the upper panels of Fig. 6 indicate 
the $68\%$ confidence limits of the best-fit profile of 
Allen, Schimdt \& Fabian (2001a). 
The profile 
was originally scaled in units of $T_{2500}$ and $r_{2500}$, here it 
has been rescaled in units of $T_{200}$ and $r_{200}$. For the considered
sample of clusters $<T_{2500}>/<T_{200}>\simeq 1.31$ and 
$<r_{200}>/<r_{2500}>\simeq 2.78$, with small dispersions.
The best-fit profile exhibits a decline in the cluster inner region and
an approximate isothermality between $r\simeq 0.1 r_{200}$ and $r \simeq
0.3 r_{200}$. From Fig. 6 it can be seen that the simulated 
profiles are unable to follow this behavior. They have a steep rise 
toward the cluster center, with peak values located at 
$r \simeq 0.02r_{200}$. Moreover they decline outward with radius, whereas the
best-fit profile stays approximately constant.
However it must be stressed that the sample  of Allen et al. is somewhat
peculiar in this aspect. Declining temperature profiles have been 
recently measured by a number of authors \cite{vi05,pi05}, using spatially
resolved spectra obtained from {\it Chandra} or {\it XMM-Newton} satellites.
The rescaled profiles are quite similar, with a decline in temperature
from its peak value at $r \simeq 0.1r_{200}$ toward outer radii. A possible 
explanation for this discrepancy is suggested by analyzing the differences 
between the profiles of different subsamples.
From the top-right panel of Fig. 6  it can be seen that the average
profile of cool clusters is steeper than that of hot clusters. This 
dependence of the shape of the profile on the cluster mass is a 
consequence of the scale dependence introduced by cooling. 
According to the results shown in the previous section the rescaled central 
\tps are expected to be higher for cool clusters than for hot clusters.
As a consequence, the slopes of the temperature profile at the cluster center
will also be higher. {This dependency of the slopes on the cluster mass 
is more clearly illustrated in Fig. \ref{fig:gradt} and is
observationally confirmed by Fig. 16 of Vikhlinin et al. (2006).
In this paper the authors analyze the gas and mass-density profiles of 
a sample of 13 low-redshift regular clusters. For this sample spectral temperature 
profiles were extracted in Vikhlinin et al. (2005), and in the lower panels of 
Fig. 6 their best-fit profile is compared against the profiles from the cooling runs 
(see later).
Fig. 16 of Vikhlinin et al. (2006) shows that the observed profiles for cool 
clusters are significantly different and steeper than for massive clusters, thus
confirming the scale dependency found here in the scaled profiles.}

The average profile of hot clusters is the one which is closest 
to the strip which is defined by the best-fit profile of 
Allen, Schimdt \& Fabian (2001a). 
The agreement is significantly improved if one considers the profiles
of active clusters in the left panel. In fact, the average profile 
of the subsample defined by both hot and active clusters is in good 
agreement with data, with the expection of the two innermost bins.
These results suggest that a proper
comparison of simulated profiles with the best-fit profile of 
Allen et al. (2001a)  must take into account the mass and the morphological 
composition of the data sample. 
From Table 3 (\La) of Allen et al. there are only two clusters 
with $T_{2500} \simeq 6 keV$, whereas $T_{mw}^{2500} \simgt 10 keV$ for the
remaining four clusters.  Here the subsample of hot and active clusters
has 5 members with $T_{mw}^{2500}$ between $5$ and $6 keV$. 
This implies that the scale dependency of the average profiles because 
of cooling is an important factor to be considered. Therefore in order 
to perform a careful comparison of the simulated profiles with the 
data of Allen et al. a sample of clusters extracted from a
simulation volume much larger than the ones used here ($\simeq 1200 Mpc$)
is needed.

Another issue concerns the role of merging events that act on the shaping of
 the cluster temperature profile. The best accord with data is obtained 
for hot clusters which are also active,  however the sample of Allen et al. 
consists of six clusters for which lensing mass measurements are in agreement
with X-ray masses. A result which is indicative of relaxed configurations. 
It is worth noticing, however, that some sort of merging activity must have 
been at work, for at least some clusters of the sample. This is indicated by 
the shape of best-fit profile, which is flat or decreases towards smaller radii.
This is at variance with the scaled profiles as measured from other samples
\cite{vi05,pi05}, which have a peak at $\simeq0.1 r_{200}$ and a decline 
outward.
A flat profile is however what is found when  moving from quiescent to active 
clusters: because of the mixing effects associated with the mergers the 
profiles have their peak height reduced or erased completely in the case
of hot clusters.
This behavior agrees with the findings of 
De Grandi \& Molendi (2002), who have investigated
 the temperature profiles for two samples of 11 non-cooling flow clusters
 (NCF) and 10 cooling flow clusters (CF). 
For NCF clusters the authors derive profiles with a core which is 
approximately
 isothermal, which in the standard scenario is interpreted as a
consequence of mergers.

In the lower panels of Fig. 6 the simulated profiles
scaled in units of $T_X$ are shown. 
The meaning of the symbols is the same as in the upper panels, the profiles
have been obtained by averaging over cluster samples at redshifts 
$z=0.116, 0.052, 0.039$ and $z=0.025$. The solid lines indicate the 
scatter ($\simeq 20\%$) in the best-fit profile to the data of 
Fig. 18 of Vikhlinin et al. (2005), which is a sample of 13 low-redshift regular clusters.
Spectral fits have been performed by Vikhlinin et al. in the energy band 
$[0.6-10]keV$, whereas the fits here are in the energy band $[0.5-7]keV$.
The differences in the spectral fits can be considered negligible, since 
all the simulated clusters have $T_{mw}^{200} \simlt 6 keV$.
The radial dependence of the best-fit profile is given by  Eq. 2 of 
Vikhlinin et al. (2005):

\be
 T/T_X= \left \{ \begin{array}{ll}
  1.07 & \mbox{if $ 0.035 <  r/r_{X} < 0.125$ } \\
 1.22  -1.2 r/r_{X}  & \mbox{if $ 0.125 < r/r_{X}< 0.6$ ~,} 
\end{array}
\right .
\label{prft}
\ee
where $r_{X}=1.95 h ^{-1} (T_X/10keV)^{1/2}$.
The solid lines have been extrapolated in the plots downward of 
$\rs < 0.035$ for the sake of a comparison with the spectral points.
For $\rs > 0.1$ the profiles are marginally consistent with data, 
at $\rs \simeq 0.2$ 
the simulated profiles are about $\simeq 10\% -20\%$ lower than the central
 value of the best-fit profile ($T/T_X\simeq 0.1$).
There is a formal improvement at $\rs <0.1$, where the flatness of the solid
lines accounts in cool cores for the scatter of the measured profiles.
Clearly, a larger sample of measured profiles of cooling flow clusters is 
needed before it would be possible to draw statistical meaningful 
conclusions about the consistency of the profiles of simulated
cooling clusters with data. 
From Fig. 15 of Vikhlinin et al. (2005), but see also Fig. 2 of 
Piffaretti et al. (2005),
there is a clear concordance that the measured scaled profiles
reach their peak values at $\rs \simeq 0.1$, whereas the simulated profiles have their peak much closer to the cluster center, at approximately 
$\rs \simeq 0.02$.  The disagreement with data  can be improved by taking
advantage of the scale dependencies previously discussed and by 
constructing a cluster subsample of a given mass and morphological composition.
However, the general framework predicts a strong correspondence between 
the presence of a cooling flow and a regular cluster morphology 
(Bauer et al. 2005, but for an alternative view see Motl et al. (2004))

\begin{figure}[t]
\begin{center}
{\includegraphics*[height=14cm,width=16cm]{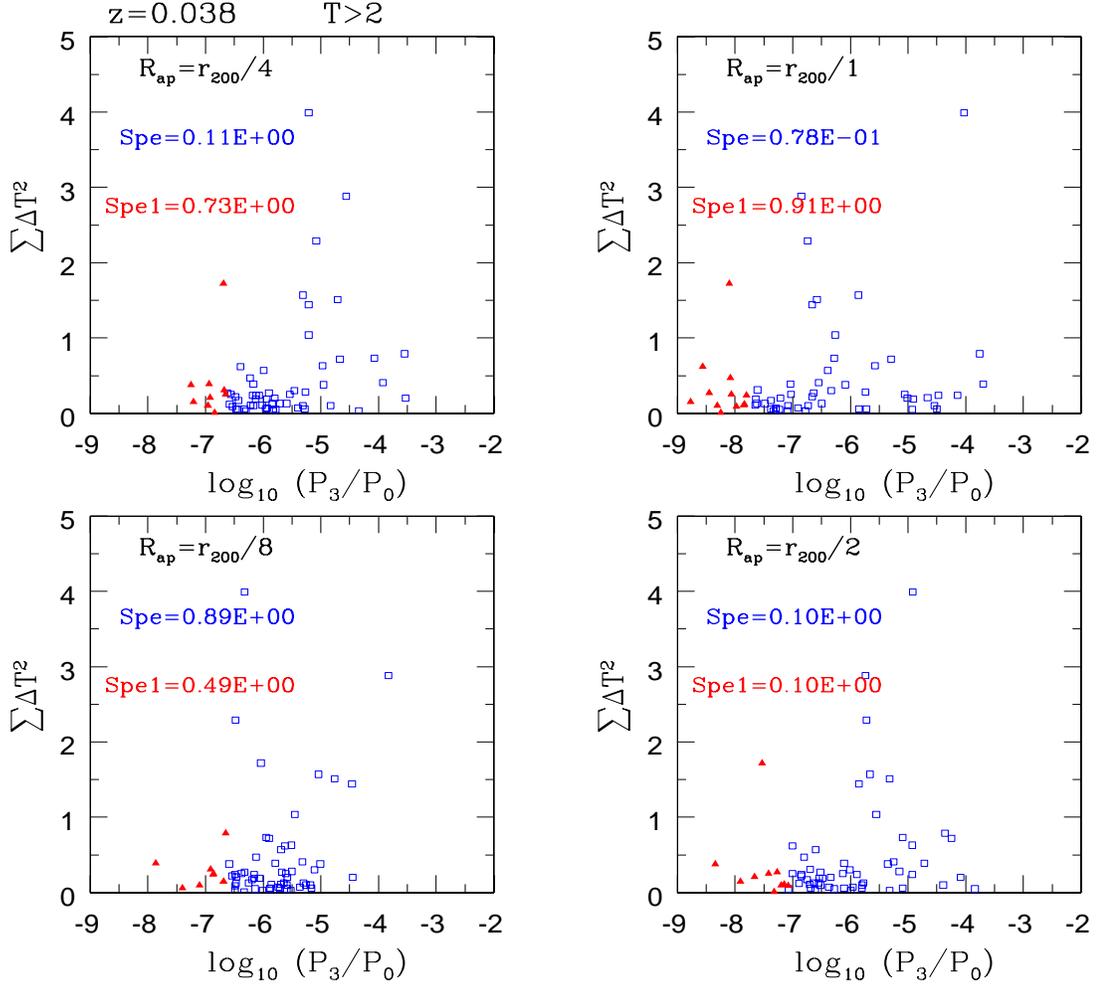}}
\end{center}
\caption{  
The sums $\sum \Delta T^2$ versus the cluster values $\Pi_3$. For a 
cluster $\sum \Delta T^2$ is defined as the sum over the radial bins, 
out to $r/r_{200}=0.1$, of the squared difference 
$\Delta T^2=[(T_S(r)-T_{ew}(r))/T_{mw}^{200})]^2$. The clusters are those
 of the sample $T>2$ of the upper panels of Fig. 6. 
Each panel is for different values of $R_{ap}$ with which  
the power ratios are calculated. Open squares (filled triangles) refer to 
all ($25\%$ percentile) of the power ratio distribution. The notation 
Spe= (Spe1=) gives the Spearman significance levels of the correlations.}
\label{fig:hisa}
\end{figure}
Because the sample of  Vikhlinin et al. consists of 11 relaxed clusters 
this indicates that one should compare with data the simulated profiles 
of the bottom right panel, rather than the profiles of active clusters
of the left panel, though a proper comparison is not possible without 
a quantitative morphological measure of the data clusters.
Although appears unlikely that the simulated temperature profiles 
of simulated cooling clusters can consistently fit the profiles 
of cooling flow clusters, note that a temperature drop at the cluster center 
is however qualitatively reproduced.

In Fig. \ref{fig:hisa} the sums 
$\sum_i \Delta T_i^2$ are shown as a function of the cluster power ratios, 
each panel is for a different value of the aperture radius $R_{ap}$.
For a given cluster the sum $\sum_i \Delta T_i^2$ is defined as the sum  
over the radial bins, out to $r/r_{200}=0.1$, of
$\Delta T_i^2=[(T_S(r_i)-T_{ew}(r_i))/T_{mw}^{200})]^2$. 
The sample investigated is the $T>2$ in the upper panels of Fig. 6.
In each panel 
 the Spearman significance levels of the correlations are given. 
The main result  is that there is a significant correlation between 
 $\sum_i \Delta T_i^2$ and $\Pi_3$, a
result which indicates that the role of merging events in reshaping
the temperature profiles can not be considered marginal.
Moreover, for quiescent clusters the correlation between 
 $\sum_i \Delta T_i^2$ and the cluster dynamical state appears 
more robust.

\begin{figure}[t]
\begin{center}
{\includegraphics*[height=8cm,width=14cm]{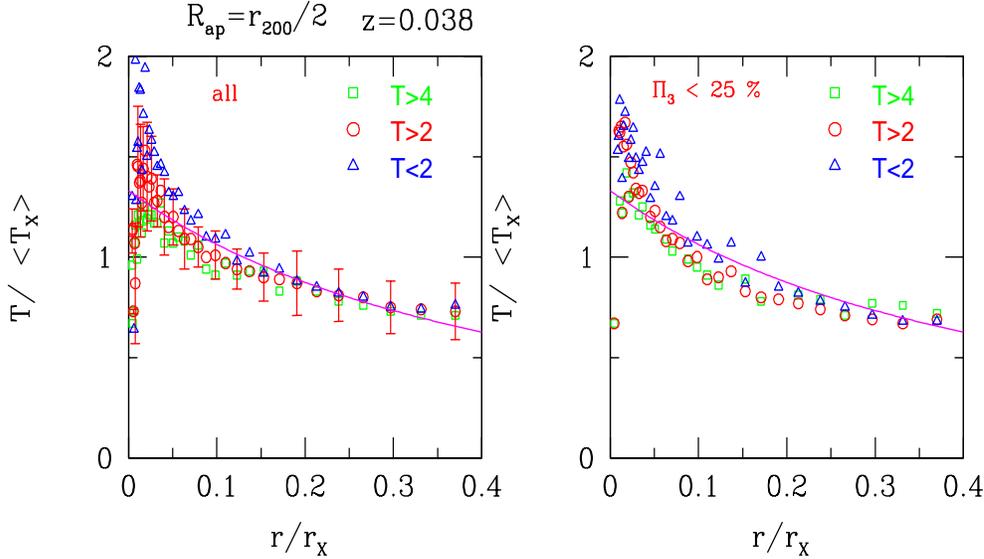}}
\end{center}
\caption{ {Average \tew profiles  are displayed for different subsamples of the 
simulated clusters at $z=0.038$. 
The notation $T>2$ indicates, as in Fig. 6, that 
all the clusters with $T_{mw}^{200}>2keV$ are part of the subsample, 
while $\Pi_3 < 25\%$ means that these clusters have a value of 
$\Pi_3$ below the
threshold which defines the $25\%$ of the cumulative distribution.}
$T_X$ is the \tew cluster temperature in the $[0.5-10]keV$ energy band
and $r_X= 1.95 h^{-1}(T_X/10keV)^{1/2} Mpc$. Continuous line is the
 best-fit
profile of Loken et al. (2002). Left panel is for all of the 
clusters while the right panel is for the cluster 
subsample $\Pi_3 < 25 \%$, the power ratios have been calculated 
setting $R_{ap}=r_{200}/2$.
}
\label{fig:tewp}
\end{figure}

\begin{figure}[t]
\begin{center}
{\includegraphics*[height=12cm,width=16cm]{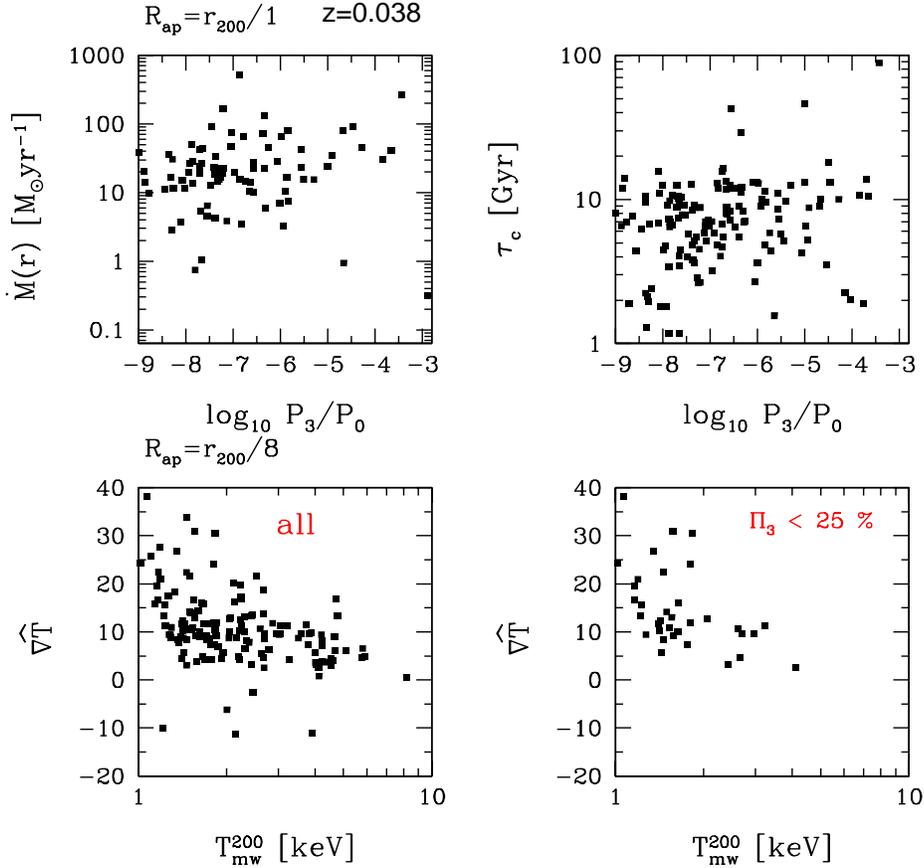}}
\end{center}
\caption{  
The mass accretion rates and central cooling times of the simulated 
clusters are displayed at $z=0.038$ in the upper panels against 
the power ratios $\log_{10} P_3/P_0$, these have been evaluated 
at $R_{ap}=r_{200}$. In the lower panels values 
at $r=0.05 r_X$ of the rescaled temperature gradient
$ \widehat {\nabla T }\equiv d (T/T_X)/d(r/r_X) $ of Fig. \ref{fig:tewp}
are plotted versus $T_{mw}^{200}$.
Left (right) panel is for all ($25\%$ percentile) of the power 
ratio distribution, with $R_{ap}=r_{200}/8$ for the power ratios.
The Spearman significance levels of the linear 
correlation coefficients is below $95\%$ only for the top left panel.
}
\label{fig:gradt}
\end{figure}

In order to assess the consistency of the simulated profiles with previous
results, for comparative purposes the work by 
Loken et al. (2002) was chosen, 
who used an adaptative Eulerian mesh hydrodynamic code. 
The projected emission-weighted profiles at $z=0.038$ are shown in
 Fig. \ref{fig:tewp}; the radial coordinate is now scaled to $r_X$ and the 
horizontal scale it extends out to $r/r_X=0.4$.
The right panel is for the subsample with $\Pi_3$ below the $25\%$ 
percentile, left panel is for all of the clusters. The solid line indicates the 
best-fit profile of Loken et al. (2002), 

\be
 T/T_X=  T_0 (1+r/a)^{-\delta},
\label{prflo}
\ee
where $T_0=1.33$, $a=r_{vir}/1.5$ and $\delta=1.6$.
Their emission-weighted profiles were constructed assuming a constant 
metallicity and using a $[1.5-11]keV$ bandpass, thus it is appropriate 
to compare the profile of equation (\ref{prflo}) with simulated profiles of 
hot or normal ($T>2$) clusters. The profiles of Fig. \ref{fig:tewp} are 
in good agreement with the best-fit profile (\ref{prflo}) over the radial 
range $0.1 \leq r/r_X \leq 0.4$, with $r/r_X=0.04$ being the lower 
limit to the fit. 
The profiles of the whole cluster sample are shallower than those of 
quiescent clusters and in better agreement with the profile of 
Loken et al. (2002).
A result which indicates that the total sample is more representative
of the sample of Loken et al. over which the fit was performed.

A comparison of the simulated spectral temperature 
profiles presented here with the data of 
Piffaretti et al. (2005) is not directly 
possible because the authors have measured their profiles using the 
cameras on board {\it XMM-Newton}. Nonetheless an indirect comparison
is still possible through their Fig. 6. In this figure the authors report, 
together with their data points, the best-fit of equation (\ref{prflo}).
The results indicate a substantial agreement of the simulated profiles with 
data out to $r/r_X \simeq 0.4$.
At small radii ($r \leq 0.05 r_X$) a comparison of the simulated profiles 
with that of Loken et al. (2002) is not possible due to the radial range of the 
fit, though it would have been interesting because of the 
different numerical methods employed in the simulations.
If one considers previous discussions and the profiles of Fig. 6
and \ref{fig:tewp} over their entire radial range, it turns out that the 
temperature profiles of cooling clusters is scale dependent and 
can not be considered universal, 
the profile of hot clusters being shallower than that of cool clusters.
This is clearly illustrated in Fig.  \ref{fig:gradt} (lower panels), where 
the dimensionless gradients $ \widehat {\nabla T }\equiv d (T/T_X)/d(r/r_X)$ 
are evaluated at $r=0.05r_X$ for the sample of Fig. \ref{fig:tewp} and the 
cluster values are shown against $T_{mw}^{200}$. The correlation of 
 $ \widehat {\nabla T }$ against $T_{mw}^{200}$ is well defined, the 
significance levels of the linear correlations are found to be above $95\%$.
This scale dependency has been found here clearly because of the large size
of the numerical  sample used; this has allowed us to subdivide the sample 
into subsamples according to several cluster properties, while keeping 
the subsamples with a statistically significant number of members.

Finally, in the top panels of Fig. \ref{fig:gradt} the 
dependencies of the mass accretion rates $\dot M$ and the central cooling
times $\tau_c$ versus the cluster dynamical state as indicated by 
$\Pi_3$ are investigated. Note that here the power ratios have been 
evaluated at 
$R_{ap}=r_{200}$, while $R_{ap}=r_{200}/8$ for the power ratios of the 
lower panels. For a given radial bin the mass accretion rate is defined as 
the spherical average of $\dot {M(r)}=4 \pi \rho_g r^2 v_r$, where 
$v_r $ is  the radial velocity of the gas. Both 
$\dot M$ and $\tau_c$ are evaluated at a fixed radius of $50kpc$.
According to the standard scenario \cite{fa94} cool gas at the  cluster 
center
will form a cooling flow as the cooling time of the gas will be shorter 
than the Hubble age. Spectroscopically the mass deposition rates of cooling 
flow clusters are found to lie in the range $\sim 10-500 M_{\odot} yr^{-1}$.
In order to mantain a steady state flow the inner region of the 
cluster must remain undisturbed by merging events. Therefore one expects 
to see some degree of correlation existing between the central cooling time 
$\tau_c$ , or the mass deposition rate $\dot M$, versus a morphological 
measure as $\Pi_3$. From the top left panel of Fig. \ref{fig:gradt}  it can be 
seen that the accretion rates $\dot M$ have a widely scattered range of
values, without displaying any clear dependence on  $\Pi_3$.
The cause of the scatter is possibly due to the adopted definition
of $\dot M$, which requires an estimate of the infall velocity  which is
subject to noise because of the local gas motion.

The cooling times $\tau_c$ show instead a significant correlation
with the values of $\Pi_3$ and thus with the cluster morphology.
The Spearman significance level  of the correlation is $p_S=0.03$. 
For quiescent clusters of the sample the average of the cooling times  gives
$<\tau_c>_q=3.75 Gyr$ and for active clusters  $<\tau_c>_a=43 Gyr$.
A Student t-test applied to the two distributions  yields $p_t=0.08$, so 
that  
the significance level for different means is above $90\%$.
This is an important result because it confirms through numerical simulations
the association between cluster morphologies and the strength of cooling 
cores. 
The plot can be compared with Fig. 8 of 
Bauer et al. (2005), who have
measured from {\it Chandra} temperature maps the cooling times and the 
power ratios of 38 
X-ray luminous clusters. The correlation of cooling times with $\Pi_3$ 
is qualitatively reproduced, though  with several differences.
The range of values of the simulated power ratios is higher than 
for data, the latter being $\simlt 10^{-6}$. Moreover, the distribution 
of the central cooling times is wider for data than for the simulations, 
with $\tau_c \simgt 10 Gyr$ for $\Pi_3 \simgt 10^{-7}$.
The main source of these differences are the different aperture radii
used by Bauer et al. ($R_{ap}=0.5 (h/0.7)^{-1} Mpc$) and the  definition
of central cooling time which is measured at the cluster innermost radial bin, 
and  is distance dependent, whereas here a fixed radius of 
$50kpc$ was used. 

A systematic analysis of the evolution of cool cores in galaxy clusters 
and how they are correlated with the cluster merger histories will be 
investigated in a future paper; here the results  on the cooling times 
are used to confirm that the temperature profiles of quiescent clusters in 
Figures 6 and \ref{fig:tewp} can be identified with  those predicted 
by the simulations for cooling flow clusters.

\section{Numerical issues}  \label {sec:numissues}
In this section it is investigated the dependence of the results presented in
this paper
on the numerical resolution of the simulations.
The latter is mainly controlled by two numerical input parameters: the  
 number of simulation particles and the chosen values 
for the gravitational softening parameters.
The most important numerical effect which must be kept under control 
in numerical simulations is the 2-body heating time $\tau_r(r)$, which must be 
at least of the order of the age of the structure and is defined as
\cite{bi87}

\be 
\tau_r \simeq 6.7 \cdot 10^5 Gyr  \left ( \frac {\sigma_1 }{10^3 Km sec^{-1}} 
\right )^3 
   \frac { h^{-2} } { (m_d /10^{11} M_{\odot})} \frac {1} {(\rho_d / \rho_c) 
\ln \Lambda }~,
\label{eq:tr}
\ee
where $\sigma_1$ is the 1-D dark matter velocity dispersion, $G$ is the 
gravitational constant, $\rho_d$ is the dark matter density; $\ln \Lambda$ 
is the Coulomb logarithm associated to the gravitational interaction,
and typical values are $ \ln \Lambda \simeq 3$;
The relaxation time $\tau_r$ can be estimated  as follows:
for a Navarro, Frenk \& White (1995) dark matter density profile \cite{lo01a}
at $z=0$ the 1-D central velocity dispersion scales as  
$\sigma \simeq 950$km sec$^{-1} (M_{vir}/10^{15} \msu)^{1/3}$.
Moreover, with the simulation method described in sect. 2, the dark particle 
mass $m_d$ of the simulated cluster can be expressed as a 
function of the mass of the cube of size $L_c$ and thus it scales linearly with 
 $M_{vir}$. 
For a lattice with  $N_L=51^3$ grid points  
it is then easy to find that $m_d \simeq 2 \cdot 10^{-5} M_{vir}$.
Accordingly, eq. \ref{eq:tr} yields:
\begin{equation}
\tau_r \simeq 40 {\rm Gyr}  
\frac {1}{ ( \rho_d/\rho_c )/10^5 \ln \Lambda }~.
\label{eq:tau2}
\end{equation}

\begin{table}[t]
\label{tab:resol}
\begin{center}
\caption{Reference values for the four 
clusters used in the numerical tests. 
$M_{200}$: cluster mass at $z=0$ within $r_{200}$ in $h^{-1} M_{\odot}$, 
$r_{200}$ in units of Mpc and $T^{200}_{mw}$ in keV. The power ratio $\Pi_3$ is 
evaluated at $R_{ap}=r_{200}$.
$\varepsilon_g$: value of the gas softening parameter in kpc,  
$N_g$: number of gas particles. These values are for the high resolution runs.
The last two columns give the values of the stellar mass 
fraction $f_{star}(<r)=M_{star}(<r)/M_{tot}(<r)$ evaluated at $r=0.1r_{200}$ 
and in units of $\Omega_b/\Omega_m$, for both the stantard (S) and the high
high-resolution (H) runs.
}
\begin{tabular}{ccccccccc}
cluster& $M_{200}$ & $r_{200}$ & $T_{mw}^{200}$ & $\Pi_3$ & $\varepsilon_g$ & 
$N_{g}$ & $f_s(S)$ & $f_s(H)$ \\
\hline
 003 &  $6.9\cdot 10^{14}$ & 2.06 & 5.26 & -6.45 & 17  &   212014 &0.108 & 0.108  \\
 008 &  $6.2\cdot 10^{14}$ & 1.99 & 4.6 & -7.32 & 17  &   212051 & 0.104 & 0.123  \\
 092 &  $7\cdot 10^{13}$ & 0.96 & 1.29 & -7.17 & 10  &   211989  &0.133 & 0.127\\
 144 &  $3.3\cdot 10^{13}$ & 0.75 & 0.95 & -7.15 & 10  &   211936 & 0.159& 0.168  \\
\hline
\end{tabular}
\end{center}
\end{table}

The heating time is then only weakly dependent on the 
 cluster mass $M_{vir}$ through the dark matter density $\rho_d(r)$. 
For a given cluster the heating time decreases as the density gets higher, 
 $\tau_r$ is evaluated at the core radius $r_c \simeq 0.01 r_{200}$, 
approximately the resolution limit of our simulations. 
For the most massive cluster of the numerical sample 
 $r_{200} \simeq 2 Mpc$, $\varepsilon_g=25 \, $kpc, 
$\ln \Lambda\sim2.6$ and $\rho_d/\rho_c \simeq 10^5$ at the radius $r_c$.
The 2--body heating time is then $\tau_r \simeq 15$ Gyr.
Similar values are obtained for the least massive clusters of the 
sample. This shows that, outside of the core radius, the simulations 
can be considered  free of 2--body  heating.

Another timescale which is relevant  
for these simulations is the cooling time 
$\tau_c =  3 n k_B T /2 \Lambda_c $,  
where  $k_B$ is the Boltzmann constant and $n$ is the gas number density.
According to Steinmetz \& White (1997) gas cooling will be 
affected by artificial 2-body heating  unless  the dark particle mass is smaller 
than the critical value 
\be 
M_c =  2 \cdot 10^9   T_6  f_{0.05} M_{\odot},
\label{eq:mc}
\ee
where  $T_6$ is the the gas temperature in units of $10^6 \gr$, $f_{0.05}$ is 
the ratio $f=\rho_g / \rho_d$ in units of $0.05$. 
For the simulated clusters studied here 
 $T_6$ lies in the range  $ T_6 \simeq 10-80$ and
$f_{0.05}\sim 2$   for $ r \simlt 0.1 r_{200}$.
Thus the critical mass 
 $M_c$ is between $\sim 3 \cdot 10^{11} M_{\odot}$ and 
$\sim 4 \cdot 10^{10} M_{\odot}$. This range of values is always above the
value of $m_d$ of the corresponding cluster by a factor $\sim 10$, so that the gas 
behavior can be 
considered free from numerical effects.

\begin{figure}[t]
\begin{center}
{\includegraphics*[height=6cm,width=16cm]{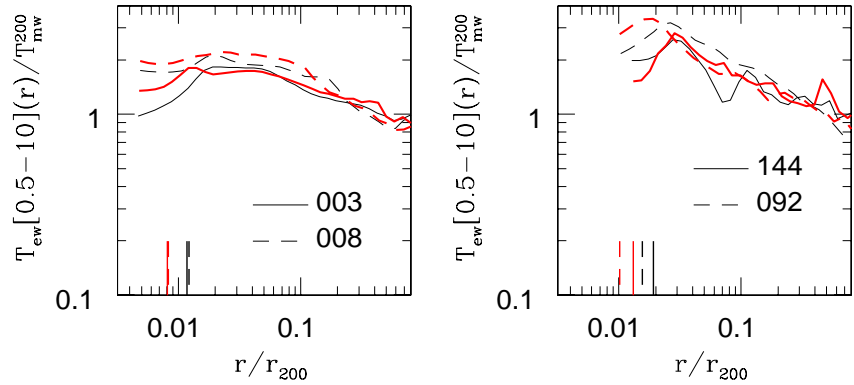}}
\end{center}
\caption{  
Radial dependence at $z=0$ of the temperature profiles  
 of the four test clusters (see Table 3).
Each cluster is labeled by its sample index, the standard simulations (sect. 2) 
are indicated by thin lines and the high resolution runs by thick (red) lines.
Left panels are for the two most massive clusters and right panels for the two
least massive. In the two panels are displayed the 
 projected emission weighted temperature profiles in the $[0.5-10]keV$ band 
in units of $T_{mw}^{200}$, the vertical lines indicate the value of 
$\varepsilon_g$ for the simulation.
}
\label{fig:numt}
\end{figure}

In order to support these conclusions, the stability of the final results
against the numerical resolution of the simulations was tested by
running again four different cluster simulations, but with a number of
particles increased by a factor $\sim 3$. The clusters were chosen in two pairs,
with the constraint of being the clusters of the first pair among the
most massive of the sample and those of the second pair among the 
least massive. The choice of the clusters was further dictated by the 
requirement of being in a relaxed state, therefore only those clusters
with a value of the power ratio $\Pi_3$, evaluated    
 at $z=0$ and at the aperture radius $R_{ap}=r_{200}$, below the threshold value
which defines the $50\%$ of the cumulative distribution were eligible.
The four clusters identified according to these criteria are listed 
in Table 3, they are identified by their index in the sample and
have $N_g \sim 210,000$. The gravitational softening parameter for
the gas particles is set to $17(10)kpc$ for the most (least) massive 
pair of clusters.
The other numerical parameters have their
values scaled accordingly. 
The stability of the final temperature profiles against the numerical
resolution of the simulation can be estimated from Fig. \ref{fig:numt}, in which 
the projected emission weighted temperature profiles in the $[0.5-10]keV$ band 
are displayed in the top panels for the
standard resolution and the high resolution runs.
For all of the test clusters there is a good agreement between the profiles 
with different resolutions,
indicating that convergence of hydrodynamical variables 
 is achieved for $N_g \simgt 70,000$.
For the less luminous clusters the similarity of the temperature profiles 
between runs with different resolutions shows that the peak location 
of the profiles is adequately resolved in the standard resolution runs
and is not adversely affected by the value of $\varepsilon_g$. 

Another numerical issue is the effect of numerical resolution on the final amount of
cooled gas turned into stars. 
In the last two columns of Table 3 for each cluster the values of 
 the star mass fraction $f_{star}(<r)=M_{star}(<r)/M_{tot}(<r)$, evaluated 
at $r=0.1r_{200}$, are given in correspondence of the standard and high-resolution 
runs, respectively.
The similarity of the values of $f_{star}$ between runs with different resolution
suggests that 
convergence is being achieved for $f_{star}$ and that the SN feedback model
implemented here effectively regulates the amount of cooled gas.
These conclusions agree with previous findings of V03 (sect. 4.4), in which it was
shown the stability of $f_{star}$ against the numerical resolution of the
hydrodynamic cluster simulations performed there. 
Finally it is worth noting that cluster X-ray properties are largely 
determined by the high-temperature part of the gas distribution, and
are not very sensitive to the amount of cooled gas in the cluster inner regions
(V03).

\section{Conclusions} \label {sec:conclusions}
In this paper results from a large set of hydrodynamic/N-body simulations 
of cooling clusters have been used to investigate the relationships 
between spectral \tps and mass-weighted or emission-weighted temperatures.
X-ray spectral \tps have been constructed according to a set of 
procedures in order to reproduce spectroscopic measurements as expected from 
{\it Chandra} satellite.
Spectral fits have been performed using the single-temperature {\it mekal} model
 of XSPEC, with three free parameters: the gas temperature, the 
metallicity abundance and the normalization.
The main difference with respect to previous works which have investigated 
these issues is in the physical modeling of the gas, which is allowed 
to cool radiatively and is subject to feedback heating by SNe.
This in turn implies a scale dependency of cluster X-ray properties 
because of the cooling efficiency to convert cold gas into stars, 
which is higher for cool clusters.
For instance, the relationships between the global spectral temperatures
$T_S^{\Delta}$ and the mass-weighted temperatures $T_{mw}^{\Delta}$  are 
found to be significantly biased with respect to those obtained from previous 
runs (ME), in which the modeling of the gas was adiabatic.
The scale dependency introduced by cooling has a significant impact also on the
value of $T_{mw}^{\Delta}$ as different cluster overdensities are considered,
therefore the $T_S^{\Delta}-T_{mw}^{\Delta}$ relation is found to be strongly 
affected  also by the value of $\Delta$. 
Moreover, the scale dependency due to cooling also enters indirectly in the 
determination of the shape of the  $T_S^{\Delta}-T_{mw}^{\Delta}$ relation
because of selection effects which limit the sample construction and 
are dependent on the cluster mass.
The biasing of spectral \tps with respect to emission-weighted temperatures
is found to be similar to that with mass-weighted temperatures, but with a 
wider scatter.

Another effect which governs the degree of biasing between spectral and
mass-weighted \tps is given by merging processes. Spectral temperatures,
as obtained by fitting single-temperature models, can be significantly biased
toward lower values than mass-weighted \tps because of accretion of 
subclumps due to hierarchical clustering which contain cooler gas that
modifies the photon spectrum. This source of biasing was discussed in detail 
by ME, here it has been analyzed using the power ratios as indicators of the 
cluster substructure. 
The main advantage of the method is the possibility to quantify the degree 
of substructure on a chosen scale, identified by the aperture radius $R_{ap}$.
This scale dependency has been investigated by evaluating for the considered 
samples the cluster power ratio $\Pi_3$  at the aperture radius 
$R_{ap}\simeq R_{\Delta}$, in correspondence to $T_S^{\Delta}/T_{mw}^{\Delta}$ 
for the cluster under consideration.
The results confirm the existence of a strong correlation between the ratio
 $T_S^{\Delta}/T_{mw}^{\Delta}$ and the cluster substructure as measured 
by $\Pi_3$. This correlation is found with a high significance level for 
all of the considered overdensities. An important result which follows
from this analysis is that the biasing associated with merging 
events is independent from the one which follows because of the scale 
dependency due to cooling.
Thus, the biasing of spectral \tps in galaxy clusters is governed by 
two processes, independent of each other, the first being due to subclump
accretion of cool gas and the second because of cooling.

This two parameter model can explain also the shape of the simulated 
temperature 
profiles. The temperature profile is however not universal even for 
relaxed clusters. This follows because, owing to the scale dependency 
introduced by cooling, the central temperature in relative units 
 is higher  for cool clusters than for hot clusters. This in turn implies 
for the former steeper profiles at the cluster center  than for 
massive clusters. The gradients of the rescaled profiles are found to 
be dependent on $T_{mw}^{\Delta}$ with a high statistical significance.
This scale dependency of the shape of the profiles on the cluster mass 
is softened or erased entirely if the cluster has undergone a number of 
major merger events. The profiles of these clusters are much shallower than 
those of quiescent clusters. This is clearly due to the effects of 
remixing of the gas due to mergers and these differences have been quantified
according to the cluster morphology using the power ratios. It has been 
found that the profiles are modified in a significant way only in the 
presence of major mergers, where the mass of the accreting subclump is 
a large fraction of the cluster mass.
These results were already suggested by Allen et al. (2001b, cf. sect. 9.2), 
but here they are confirmed through numerical simulations.

For quiescent clusters the scaled temperature profiles are in good agreement
with the considered data in the radial range $0.1 \simlt r/r_X\simlt 0.4$. 
The profiles are not consistent with those measured from a sample of massive
clusters ( Allen et al. 2001a), but the disagreement can be made less severe 
if the simulated sample is constructed by taking into account the scale 
dependencies due to cooling and merging events. 
A more severe discrepancy is instead found at small radii ($r/r_X \simlt 0.1$)
between the measured profiles and those of the simulations. The profiles 
of the simulations reach their peak values at the scaled radial coordinate
 $r/r_X \simeq 0.02$, which is smaller by a factor $\sim 5$ of the peak
location as derived by observations.
Moreover, the peak height of the simulated temperature profiles is $\sim 20 \%$ 
higher than that measured. These discrepancies are clearly indicative that 
the cooling model fails to consistently reproduce  the observed temperature
profiles. In a recent paper Motl et al. (2004)  argue that hierarchical merging
can provide a viable scenario for the formation and evolution of cool cores 
in galaxy clusters. 
The results of Fig. 6 indicate that merging events play a key role 
in reshaping the cluster profiles and support the proposed model. However, 
it is worth noticing that a significant reshaping of the temperature peak
is achieved only in the regime of strong merging ($\Pi_3 \simgt 75\%$);
moreover  this regime has the side effect that the central temperature drop
is no longer reproduced.

The measured temperature profiles have a peak location in rescaled radial 
units which is located at a larger distance from the cluster center than in
 the simulations. 
This is indicative that real clusters, at some stage of their evolution, 
have undergone a significant diminution of their cooling rates
in their cores.
This is in order to provide pressure support and to prevent the inflow of 
gas at a higher entropy toward the cluster center. The suppression of the 
cooling
rate is most easily achieved if the gas is heated by some energy source. 
The most plausible candidates are SNe whose feedback energy from their 
explosions can heat the gas to the desired level. However the results 
of the simulations performed here show that this result is not 
achieved even in the case of 
maximum theoretical efficiency of transfer of energy to the ICM. {This indicates 
that the complex nature of the thermal structure of the  cluster gas
 is not adequately described by the physical processes incorporated  in the cooling 
runs performed here and suggests that additional physics is clearly needed
in order to prevent overcooling in cluster cores.
These results agree with previous findings and have prompted many authors
to consider possible physical sources of additional heating to the ICM.
The most popular models are (cf. Voit 2005 and references cited therein) energy 
feedback from active galactic nuclei (AGN) and thermal conduction. 
Because of the existing temperature gradients the latter model could in principle 
provide the necessary heating in the
cluster central regions \cite{za03,vo04},
but is currently disfavored by more sophisticated numerical 
simulations which include thermal conduction \cite{do04,po05}.

The most favored scenario is thus the AGN  model, in which 
a supermassive black hole at the center of the cluster is fueled
by the gas of the cooling flow and can balance the energy losses of the
gas through some energy transfer.
A popular mechanism involves heating of the gas due to energy dissipation
of buoyant bubbles, inflated by the AGN near the cluster center and rising
through the ICM \cite{cu01,br02,fab02,ru02}.
The dissipation can occur through a number of processes, such as viscous
dissipation or turbulence.
It is clear that incorporating these physical processes  into numerical simulations
 of cluster 
formation is a necessary step in order to 
explain the shape of the observed profiles.
This is a very demanding computational challenge which has only recently 
begun to be addressed \cite{br02,be02,ba03,om04,ru04,br05,si06}. \
In such a scenario it is difficult to assess how the main findings
presented in this paper will be modified. Although a proper 
analysis can only be performed through numerical simulations, 
several qualitative arguments can nonetheless still be used to address the effect
 of the interaction of the AGN with the ICM
on the biasing of spectral  temperatures.

In the proposed AGN heating model buoyant bubbles of plasma will rise from 
the cluster center toward the surrounding medium, where the transported energy
will be transferred into the ICM and converted to heat. This will produce
some degree of thermalization of the ICM, which in turn implies a reduction
in the amount of cold gas and thus in the soft X-ray emission, thereby 
reducing the degree of biasing of spectral \tps.
In other words, for a given cluster mass scale, any additional source of 
feedback is expected to reduce the biasing.
Another issue is if the scale dependency introduced by cooling
will be preserved in the heating scenario. Some degree of modification
will be present, since the bubble injection energy, $E_{bub}$, is 
scale dependent itself, according to the black hole mass.
 For example, in one of their models Sijacki \& Springel (2006) propose 
$E_{bub} \propto M_{200} ^{4/3}$, and therefore in  relative terms the energy 
feedback is higher for massive clusters.
The effects on the thermal status of the ICM of energy feedback from AGN are however
not expected to modify in a significant way the scale dependency due to cooling.
This argument is mainly supported by recent measurements of
spectral temperature profiles for a sample of cooling flow clusters 
(Vikhlinin et al. 2005, 2006). The radial behavior of the measured profiles 
 for cool clusters is  significantly different and steeper than for massive 
clusters, which is in qualitative agreement with what found here.
These conclusions are valid if the cluster possesses a regular X-ray morphology, 
which is indicative of a dynamically relaxed status.
In the case of an ongoing merger large amount of cold gas could be supplied at the
cluster center, thus replenishing the gas reservoir of the black hole and 
triggering AGN activity. In this scenario the thermal status of the ICM surrounding
the black hole will be due to a complex interplay between the accretion of cold
gas and the uplifted bubble energy. This suggests that the scale dependency depicted
 in Fig. \ref{fig:dtpw} will be most likely modified in the proximity of a major
merger event ($\Pi_3 \rightarrow 0$), with a  decrease in  the spectral bias owing 
to the feedback energy due to accretion and injected into the ICM.
Such a complex picture can however be addressed self-consistently only through
numerical simulations.

To summarize, it is worth noting, however, that any proposed heating model must 
be able 
to explain the lack of cold gas at the cluster center as well as the observed 
peak location of the temperature profile. 
Finally, the results of this paper indicate that in order to properly 
compare X-ray properties of simulated clusters with those observed, 
the mass and the morphological composition of the sample must be
carefully taken into account.}

\section*{Acknowledgments}
The author is grateful to K. Arnaud, A. Gardini and P.Tozzi for their
helpful comments and patient explanations which made it possible 
to develop the procedures described in sect. 3.

\end{document}